\documentclass[a4paper]{article}

\usepackage{amsmath,amsthm}      
\usepackage{graphicx}            
\usepackage{microtype}           
\usepackage{booktabs}            
\usepackage{multicol}            
\usepackage[hidelinks]{hyperref} 
\usepackage{physics}
\usepackage{mathtools}
\usepackage{relsize}
\usepackage{amssymb}
\usepackage{setspace}
\usepackage{pdfpages}
\usepackage{caption}
\usepackage{subcaption}
\usepackage[english]{babel}
\usepackage[utf8]{inputenc}
\usepackage{authblk}

\theoremstyle{definition}

\newcommand{\lbl}[1]{\label{eq: #1}}
\newcommand{\rf}[1]{(\ref{eq: #1})}

\newcommand{\prt}[1]{\partial_{#1}}
\newcommand{\inv}[1]{\frac{1}{#1}}

\newcommand{\ttx}[1]{\textit{#1}}

\newcommand{\vx}{\vb{x}}

\newcommand{\intRn}{\int_{\mathbf{R}^n}}
\newcommand{\mathsym}[1]{{}}
\newcommand{\unicode}[1]{{}}
\newcommand{\ba}{\begin{align}}
\newcommand{\ea}{\end{align}}

\begin{document}

\title{Thermodynamics of light }
\author[]{Per Kristen Jakobsen}
\affil[]{\it \small Department of Mathematics and Statistics, the Arctic University of Norway,\newline 9019 Troms\o, Norway}
\date{}
\renewcommand\Authands{ and }
\maketitle
\tableofcontents

\section{Introduction}
These notes where written as a running documentation of a project whose goal was to explore the thermodynamical properties of light in a cavity. The notes does not make claims of originality with respect to any the topics that is discussed, perhaps with the exception og 2.1.4, 2.1.5 and 5.4.2, but we believe that it is useful to collect together the basic descriptive tools needed to discuss thermal fields in cavities together with a  nontrivial example in one place with a consistent notation and point of view. We surely would have been happy  to discover these notes while we where starting up our project. The  inspiration of our project and hence these notes  was papers published on the subject by Masud Masuripur and Pin Han\cite{Masud1},\cite{Masud2}.  In addition to this introduction, the notes consists of four separate sections.

  In the second section we introduce the thermodynamical formalism that we will be using. The formalism is of course well known and has been around for at least 150 years, but we include a derivation here because we will use the opportunity to remind the readers that the foundation of the subject, which has been argued over for at least 100 years has, starting in the 1950s, largely been resolved.
   
     It has of course been known since the seminal work on kinetic theory starting in 1843 with Waterston and continuing by Helmholtz, Clausius and Maxwell, Boltzman and last but not least Gibbs, that the foundation of thermodynamics is statistical mechanics. In fact, it was Gibbs\cite{Gibbs} who lay much of the foundation for how we think about the subject today, in particular he was the first to observe that a certain variational principle is hiding at the heart of statistical mechanics.

  There has however been a long running controversy over exactly how statistical mechanics provide a justification for the thermodynamical formalism. The key problems that have been argued over is how statistical mechanics can be used to justify the assumption of thermodynamic equilibrium and the second law. The problems has been vigorously pursued by scientists of the highest caliber and the controversy has over the years spawned entire new fields of mathematics, like ergodic theory and dynamical system theory. One of the important foundational problem for thermodynamics, the equilibrium assumption,  was however not fully resolved until the full ramifications of Shannons seminal paper on a mathematical theory of communication from 1948\cite{Shannon}, was realized. Many scientists took part in the effort of founding thermodynamics on the mathematical theory of information that grew out of Shannon's 1948 paper. However, in this effort E. T. Jaynes stands out. He both initiated the effort in 1957\cite{Jaynes1} and he was also integral to the effort\cite{Jaynes2} right up until he died  in 1998,  leaving his monumental book on the subject\cite{Jaynes3} unfinished. There however exists a well written two-volume textbook on statistical mechanics by W. T. Grandy\cite{Grandy} which is using the information theoretical foundation throughout the text.
  
    A careful reader might register that I am a little guarded when I speak about the resolution of the foundational problems of thermodynamics through information theory; I say that the nature of the equilibrium assumption has been fully understood through information theory, I don't say that the whole foundational problem of thermodynamics has been resolved through information theory. This is because in my opinion the foundation of the second law of thermodynamics has not been resolved in this way. Papers by both Jaynes, Grandy and also others introduce arguments that purport to show that the second law can indeed be derived from information theory, but I find these arguments  unconvincing.
 
In the third section we describe the quantization of free electromagnetic fields in a class of cavities that will be the focus of these notes. Field quantization is in general an immensely technically demanding and subtle part of physics. However, most of these technicalities and subtleties only come into play for interacting gauge fields, like the fields defining the Standard Model of elementary particle physics. The quantization of the electromagnetic field in a cavity is by comparison technically and conceptually simple. In fact this was the first kind of field quantization that was achieved\cite{Born} and  it's basic idea of identifying the shape of photons, the quanta of the electromagnetic field, with  classical electromagnetic cavity modes is an idea that permeate the whole area of quantum field theory. We include a detailed, but simple, account of the quantization of the electromagnetic  field in cavities because it is beautiful and fundamental, but also in order to fix our notation. 

In the fourth section our focus is on the thermodynamics of the free quantum electromagnetic field in the class of cavities described in section three. The state of the field will thus always be given by the density operator for a noninteracting bosonic many particle system. The mathematical description of this density operator is  well known from the literature , but we nevertheless  include a detailed description of the density operator. The reason we do this is both to fix the notation we will be using when we study specific cavities in section five, but also because, in our humble opinion, the notation used to describe the bosonic many particle density operator is often excessive and confusing. We introduce a somewhat unusual notation for the number state basis for the bosonic Fock space which we believe is more compact and less confusing than the standard one.

In the fifth and last section we apply the formalism developed in the four first section to investigate thermodynamical properties of the free electromagnetic field in certain specific cavities.

\section{Thermodynamics}

\subsection{The maximum entropy principle for classical systems}

Let $x_1, ... , x_n$ be random variables with an associated probability distribution $\rho(x_1,...,x_n)$. Let $f_1(x_1,...,x_n),...,f_p(x_1,...,x_n)$ be functions defined on the space of random variables $\Omega=\{(x_1,x_2,...,x_n)\}$ where the variables $x_n$ can run over a finite set, an infinite discrete set, for example a set indexed by a finite set of integers, or the variables can run over the real numbers. We will usually think about the real number case and will therefore write integrals instead of sums.
The functions $f_j$ are our \ttx{observables}. Their $\it{expectation}$ values are as usual defined by 
\begin{align}
\left<f_j\right> \; = \int_{\mathbf{R}^n} dV \; f_j(x_1,...,x_n) \; \rho(x_1,...,x_n). \lbl{75} 
\end{align}
The expectation value of a given observable of course depends on which probability distribution, $\rho$, we use. The challenge in statistics is to figure out which probability distribution one should use in any given situation. Let us say that we for some reason, (expert knowledge,guesswork, hearsay, ...) believe that a probability distribution $\rho_0$  accurately represents what we currently know about a given system. The probability distribution $\rho_0$ is called the \textit{prior distribution}, or just the \textit{prior}.

Let us next assume that we measure the mean values of the observables $f_1, ...,f_p$ and find the values $c_1,...,c_p$. If 
\begin{align}
\left<f_j\right>_0 \;  = \int_{\mathbf{R}^n} dV \; f_j \; (x_1,...,x_n) \; \rho_0 (x_1,...,x_n) = c_j, \lbl{76} 
\end{align}
we are satisfied with our choice of prior. It predicts exactly the mean values that are observed. \\
But we might not be so lucky. Perhaps 
\begin{align}
\left<f_j\right>_0 \; \ne c_j, \lbl{77} 
\end{align}
for at least one $j$. Our selected $\rho_0$ is then not the correct one, it predicts expectation values that are not observed. The challenge is to modify $\rho_0$ into a new distribution $\rho$ that is consistent with \ttx{all} the observed mean values. 

For this purpose we define a functional $S(\rho)$ by 
\begin{align}
S(\rho) = - \int_{\mathbf{R}^n} dV \; \rho \; \ln (\frac{\rho}{\rho_0}). \lbl{78} 
\end{align}
$S$ is by definition the \ttx{relative entropy} of the probability distribution $\rho$ with respect  to $\rho_0$. We will see later that our use of the word entropy here is consistent with its usage in thermodynamics. 

The \ttx{maximum entropy} principle states that one should choose the probability distribution that maximizes the functional 
\begin{align}
S(\rho) = - \int_{\mathbf{R}^n} dV \; \rho \; \ln (\frac{\rho}{\rho_0}), \lbl{79} 
\end{align}
subject to the constraints 
\begin{align}
\left<f_j\right> \; = \int_{\mathbf{R}^n} dV \; f_j \; \rho = c_j,\;\; j=1,2,...,p. \lbl{80}
\end{align}
\\
\subsubsection{The general thermodynamical formalism}
In this section we will solve the maximum principle stated in the previous section using the calculus of variations. The problem will initially be solved in the general setting described in the previous section but we will eventually specialize to the case of statistical mechanics. 

In order to proceed we must first recognize that in additional to the $p$ constraints \rf{80}, we have one more constraint that simply expresses the fact that $\rho$ is a probability distribution.
\begin{align}
\left<1\right> \; = \int_{\mathbf{R}^n} dV \; \rho (x_1,...,x_n) = 1, \lbl{206} 
\end{align}
and we thus have $p+1$ constraints and therefore introduce an extended functional 
\begin{align}
T(\rho) = S(\rho) - \lambda_0 \left<1\right> - \mathlarger{\sum}^p_{j=1} \lambda_j \; \left<f_j\right>, \lbl{207} 
\end{align}
Note that we could have written 
\begin{align}
T (\rho) = S(\rho) - \lambda_0 \; ( \left<1\right> - 1) - \mathlarger{\sum}^p_{j=1} \lambda_j \; ( \left<f_j\right> - c_j), \lbl{208} 
\end{align}
in order to make the values of the constraints explicit. However, all constant terms vanish when we take variational derivative, so we might as well drop the constant terms. Also note that our choice of minus sign in front of the Lagrange multiplier terms in \rf{207} and \rf{208} is a convention inspired by the application of this formalism to the case of statistical mechanics.

\noindent The integral density corresponding to the extended functional $T (\rho)$ is 
\begin{align}
\mathcal{L} = - \rho \; \ln (\frac{\rho}{\rho_0}) - \lambda_0 \; \rho - \mathlarger{\sum}^p_{j=1} \lambda_j \; f_j \; \rho \lbl{209} 
\end{align}
Observe that $\mathcal{L}$ does not depend on any derivatives of $\rho$. The Euler-Lagrange equation for $T$ is therefore simply 
\begin{align}
\frac{\partial \mathcal{L}}{\partial \rho} &= 0, \lbl{210} \\ 
&\Updownarrow \nonumber \\
& - \ln (\frac{\rho}{\rho_0}) - 1 - \lambda_0 - \mathlarger{\sum}^p_{j=1} \lambda_j f_j=0 , \nonumber
\end{align}
whose solution is
\begin{align}
\rho &= \frac{\rho_0}{Z} \; \exp{-\mathlarger{\sum}_j \lambda_j \; f_j}, \nonumber 
\end{align}
where we have defined $Z=\exp{(1+ \lambda_0 )}$. In order for the constraint $\left<1\right> = 1$ to be satisfied, we must have 
\begin{align}
\left<1\right> &= 1, \nonumber \\
&\Updownarrow \nonumber \\  \intRn dV \; \frac{\rho_0}{Z} \; \exp{-\mathlarger{\sum}_j \lambda_j \; f_j} &= 1, \nonumber \\ 
&\Updownarrow \nonumber \\ Z = Z(\lambda_1,...,\lambda_p) &= \intRn dV \; \rho_0 \; \exp{-\mathlarger{\sum}_j \; \lambda_j \; f_j}, \lbl{211} 
\end{align}
and the stationary distribution is 
\begin{align}
\rho(x_1,...,x_n) = \frac{\rho_0(x_1,...,x_n)}{Z(\lambda_1,...,\lambda_n)} \; \exp{-\mathlarger{\sum}^p_{j=1} \lambda_j \; f_j (x_1, ...,x_n)}. \lbl{212} 
\end{align}
$\rho$ is called the \ttx{maximum entropy distribution} and $Z$ is the $partition function$. Note that we have not proved that the distribution \rf{212} in fact gives a maximum value  for $S$, but this can be done\cite{Jaynes4}. \\
The Lagrange multipliers $\lambda_1,...,\lambda_p$ are chosen so that all the constraints are satisfied 
\begin{align}
\left<f_j\right> \; = \intRn dV \; f_j (x_1,...,x_n) \; \rho(x_1,...,x_n) = c_j && j=1,...,p\;. \lbl{213} 
\end{align}
The system of equations \rf{213} consists of  $p$ equations for the $p$ unknown quantities $\lambda_j$. 

As it turns out, we almost never need to know the distribution $\rho$ from \rf{212}, it is enough to know the partition function. Observe that 
\begin{align}
\left<f_j\right> \; &= \intRn dV \; f_j \; \rho\nonumber\\ 
&= \inv{Z} \; \intRn dV \; f_j \; \rho_0 \; \exp(-\mathlarger{\sum}^p_{i=1} \lambda_{i} f_{i}) \nonumber \\ 
&= -\inv{Z} \; \intRn dV \; \partial_{\lambda_j}\{ \rho_0 \; \exp(-\mathlarger{\sum}^p_{i=1} \lambda_{i} f_{i})\} \nonumber \\ 
&= -\inv{Z} \; \partial_{\lambda_j}  \intRn dV \;  \rho_0 \; \exp(-\mathlarger{\sum}^p_{i=1} \lambda_{i} f_{i}) \nonumber \\
&= -\inv{Z} \; \prt{\lambda_j} Z =- \prt{\lambda_j} \ln Z \nonumber \\
&\Downarrow\nonumber\\
\left<f_j\right> \; &=- \prt{\lambda_j} \ln Z,\lbl{214} 
\end{align}
and thus we can find the mean of all the quantities $f_j$ by taking partial derivatives of the natural logarithm of the partition function with respect to the Lagrangian multipliers. Moreover, we also have 
\begin{align}
\prt{\lambda_j \lambda_k}\ln Z &= \prt{\lambda_j} (\inv{Z} \; \prt{\lambda_k} Z) \lbl{215} \\
&= - \inv{Z^2} \; \prt{\lambda_j} Z \; \prt{\lambda_k} Z + \inv{Z} \; \prt{\lambda_j \lambda_k} Z \nonumber \\
&= - \prt{\lambda_j} \ln Z \; \prt{\lambda_k} \ln Z + \inv{Z} \; \intRn dV \; f_j \; f_k \; \rho_0 \; \exp(-\mathlarger{\sum}_i \lambda_i \; f_i) \nonumber \\
&= - \prt{\lambda_j}\ln Z \; \prt{\lambda_k} \ln Z + \left<f_j \; f_k\right>. \nonumber 
\end{align}
Thus 
\begin{align}
\left<f_j \; f_k\right> \; = \prt{\lambda_j} \ln Z \; \prt{\lambda_k} \; \ln Z + \prt{\lambda_j \; \lambda_k} \ln Z \lbl{216} 
\end{align}
In a similar way \ttx{all} correlation coefficients $\left<f_1^{n_1} ... f_p^{n_p}\right>$ can be expressed through partial derivatives of the partition function.

Inserting the maximum entropy distribution \rf{212} into the entropy functional \rf{79} gives us the following expression for the maximal value of the entropy
\begin{equation}
S=\ln Z+\mathlarger{\sum}_j \; \lambda_j \;\left< f_j\right>.\lbl{MaxEntropy}
\end{equation}
From a mathematical point of view we now have two sets of variables $\{\left< f_1\right>,...,\left< f_p\right>\}$ and $\{\lambda_1,...,\lambda_p\}$. Geometrically we imagine that these two pairs of variables, together with $S$, defines a space $\Omega$  of odd dimension $2p+1$ with coordinates $\{S,\left< f_1\right>,...,\left< f_p\right>,\lambda_1,...,\lambda_p\}$. The $p$ identities \rf{213} defines a $p+1$ dimensional surface $\Lambda$ in $\Omega$.

\noindent Taking the differential of the identity \rf{MaxEntropy} we get
\begin{equation*}
dS=\mathlarger{\sum}_j\; \frac{\partial\ln Z}{\partial\lambda_j}d\lambda_j+\mathlarger{\sum}_j \;\{\left< f_j\right>\;d\lambda_j+\lambda_jd\left< f_j\right>\}.
\end{equation*}
Restricting this differential to the surface $\Lambda$, and thus using the identities \rf{213}, gives us the following expression for the differential $dS$ restricted to the surface $\Lambda$
\begin{equation}
dS=\mathlarger{\sum}_j \;\lambda_jd\left< f_j\right>\lbl{dS1}.
\end{equation}
The identity \rf{MaxEntropy} defines the entropy as a function depending on all $2p$ variables in $\Omega$. We therefore have
\begin{equation}
dS=\mathlarger{\sum}_j\;\frac{\partial S}{\partial\lambda_j}d\lambda_j+\mathlarger{\sum}_j\;\frac{\partial S}{\partial\left< f_j\right>}d\left< f_j\right>\lbl{dS2}.
\end{equation}
Comparing \rf{dS1} and \rf{dS2} we conclude that on the surface $\Lambda$ we must have the identities 
\begin{align}
\frac{\partial S}{\partial\lambda_j}&=0,\nonumber\\
\frac{\partial S}{\partial{\left< f_j\right>}}&=\lambda_j.\lbl{SRelation1}
\end{align}
\noindent
Thus, on the surface $\Lambda$, the entropy depends only on the variables $\{\left< f_1\right>,...,\left< f_p\right>\}$ and the derivative with respect to these variables determines the values of the Lagrange multipliers in terms of the data $\{c_1,...,c_p\}$ of the problem.

It is frequently the case that in addition to the variables $\{x_1,...,x_n\}$, the observables depends on parameters. For notational simplicity, let us assume that there is only one parameter denoted by the symbol $\alpha$. Thus we have observables $\{f_1(x_1,...x_n;\alpha),...,f_p(x_1,...x_n;\alpha)\}$. The presence of the parameter does not change the argument leading up to the maximum entropy distribution \rf{212} and thus we have the formulas
\begin{align}
\rho(x_1,...,x_n;\alpha)& = \frac{\rho_0(x_1,...,x_n)}{Z(\lambda_1,...,\lambda_p;\alpha)} \; \exp{-\mathlarger{\sum}^p_{j=1} \lambda_j \; f_j (x_1, ...,x_n;\alpha)},\nonumber\\
 Z(\lambda_1,...,\lambda_p;\alpha) &= \intRn dV \; \rho_0 \; \exp{-\mathlarger{\sum}_j \; \lambda_j \; f_j(x_1,...,x_n;\alpha)}.\lbl{parameterZ}
\end{align}
 \noindent Differentiation of the partition function \rf{parameterZ} with respect to the parameter $\alpha$ gives us the expression
 \begin{align}
 \frac{\partial Z}{\partial \alpha}&=-\intRn dV \; \rho_0 \;{\mathlarger{\sum}_j \; \lambda_j \;\frac{\partial  f_j}{\partial \alpha}} \exp{-\mathlarger{\sum}_j \; \lambda_j \; f_j},\nonumber\\
 &\Downarrow\nonumber\\
  \frac{\partial \ln Z}{\partial \alpha}&=-\mathlarger{\sum}_j \; \lambda_j \;\left<\frac{\partial  f_j}{\partial \alpha}\right>.\lbl{DParameterZ}
 \end{align}
 If we repeat the calculation leading from \rf{MaxEntropy} to \rf{dS1} for the case when the observables depends on a parameter $\alpha$, we now get instead of \rf{dS1} the following more general expression for the differential of the entropy
 \begin{equation}
dS=\mathlarger{\sum}_j \;\lambda_jd\left< f_j\right>-\mathlarger{\sum}_j \;\lambda_j \;\left<\frac{\partial  f_j}{\partial \alpha}\right>d\alpha\lbl{dS3},
\end{equation}
where we have used the identity \rf{DParameterZ}.
 
 Note that this differential identity can be written in the form
  \begin{align}
dS=\mathlarger{\sum}_j \;\lambda_jdQ_j\lbl{dS3},
\end{align}
where we have introduced the quantities $dQ_j$ representing {\it generalized heat} associated with the observables
\begin{align}
dQ_j&=d\left< f_j\right>-\left<\frac{\partial  f_j}{\partial \alpha}\right>d\alpha\nonumber\\
&=d\left< f_j\right>-\left<\frac{\partial  f_j}{\partial \alpha}d\alpha\right>\nonumber\\
&=d\left< f_j\right>-\left<df_j\right>.\lbl{GeneralizedHeat}
\end{align}
Formula \rf{GeneralizedHeat} tells us what heat actually represents. Physical systems on the human scale, these are evidently the ones of most immediate interest to us, consists of an immense number of elementary subsystems. The detailed configurational variables for all these elementary systems defines the {\it  microscopic} degrees of freedom of the human scale system. The state of these microscopic degrees of freedom are unknown to us and our ability manipulate then directly is entirely lacking. The few degrees of freedom of the system whose state we {\it  can} know and which we have the means to manipulate defines the {\it macroscopic} degrees of freedom for the system. In our description of thermodynamics these are the observables $f_j$. A change in the  mean value of a macroscopic degrees of freedom,$d\left<f_j\right>$, comes from two sources. The first source is a change in the observable representing the said macroscopic degree of freedom, this is the kind of change that we have the ability to induce by direct manipulation. This quantity is represented by $\left<df_j\right>$ in formula \rf{GeneralizedHeat}. When this quantity is subtracted from $d\left<f_j\right>$ , what remains is the second source of change of the mean. This second source  is a change in the underlying probability distribution which represents a change in our information about the microscopic state of the system. When our ignorance about the microscopic state of a system increase the system grows ``hotter'', corresponding to an increase in $dQ_j$.

As is usual in thermodynamics, the formalism is misleading in the sense that $dQ_j$ merely denote an infinitesimal amount of generalized heat and is {\it not} the differential of some function $Q_j$. No such function exists. The proper mathematical way to think about the identity \rf{dS3} is that $dS$ and $dQ_j$ are differential forms where $dS$ is an exact differential forms, meaning it is the differential of a function, and $dQ_j$ are inexact differential forms and thus not the differential of a function. However, the mathematical formalism of differential forms must be  introduced in the very large context of differential geometry and we will not digress into this area of mathematics.

The above explanation of the nature of heat referred to the original application of the thermodynamical formalism where the systems has an immense number of microscopic degrees of freedom which are in principle knowable and controllable but  in praxis not. We however know that the thermodynamical formalism can be applied to any situation where systems has more degrees of freedom than the ones we chose to observe. This might be because the underlying degrees of freedom are unknown but it could also be the case that they are known but that we for various reasons choose to ignore them. In both cases the argument above stands and the existence of the unknown or ignored degrees of freedom manifest as heat in the theory.

 \noindent We will now derive a generalized version of  identity \rf{DParameterZ} that plays a crucial role when the thermodynamical formalism is applied to the special case for which the underlying space is the state space for a physical system. The system could be a classical mechanical system consisting of a finite number of particles, a system of classical  fields or, which will be our main application, the Fock state space for a quantum mechanical many particle system.
 
 In all these cases one consider systems that are confined to a bounded spatial domain $D$ which is defined by its bounding surface $\Gamma$. Thus all observables for the system will typically depend on the bounding surface $\Gamma$, $f_j=f_j(x_1,...x_n;\Gamma)$. We will now consider a small deformation, $\delta\Gamma$, of the bounding surface $\Gamma$. Thus $\Gamma\longrightarrow\Gamma+\delta\Gamma$. This deformation leads to variations
 \begin{align}
 \delta_\Gamma f_j(x_1,...x_n;\Gamma)=f_j(x_1,...x_n;\Gamma+\delta\Gamma)-f_j(x_1,...x_n;\Gamma),\nonumber\\
 \delta_\Gamma Z(\lambda_1,...,\lambda_p; \Gamma)=Z(\lambda_1,...,\lambda_p; \Gamma+\delta\Gamma)-Z(\lambda_1,...,\lambda_p; \Gamma).\lbl{variations}
 \end{align}
 Arguing exactly like we did for the simple case of a single parameter we now find the important identity
 \begin{align}
 \delta_\Gamma \ln Z=-\mathlarger{\sum}_j \; \lambda_j \;\left< \delta_\Gamma f_j\right>.\lbl{FundamentalVariationalIdentity}
  \end{align}
  \noindent This identity will, for the special cases mentioned above, lead to the definition of the thermodynamic pressure and related quantities. 
  Corresponding to the {\it differential} identity for the entropy \rf{dS3} we now get the following more general {\it variational} identity
\begin{align}
\delta S=\mathlarger{\sum}_j \;\lambda_jd\left< f_j\right>-\mathlarger{\sum}_j \;\lambda_j \;\left< \delta_\Gamma f_j\right>\lbl{dS5}.
\end{align}

\subsubsection{The thermodynamic formalism in statistical physics} 
 
   Let us now consider the special case when our underlying space is the classical state space for a mechanical system with $n$ degrees of freedom. This could for example consist of $n$ mass points. We will assume that the system is confined to a bounded domain $D$ in $\mathbf{R}^3$ defined by a bounding surface $\Gamma$. The state space is thus a subset of the euclidean space $\mathbf{R}^{6n}$ with coordinates $(\vb{q},\vb{p})=(\vb{q}_1,...,\vb{q}_n,\vb{p}_1,...,\vb{p}_n)$,  since we need 3 position coordinates and 3 momentum coordinates for each particle in order to uniquely specify the state of the system. \\
Let $\mathcal{H} = \mathcal{H}(\vb{q},\vb{p})$ be the Hamiltonian for the system of mass  points. Recall that the value of the Hamiltonian on any given state, $(\vb{q},\vb{p})$, is the energy, $E$,  of that state. \\
When $n$ is large it is very hard, and also mostly useless, to try to track the exact state $(\vb{q}(t),\vb{p}(t))$ of a system of mass points. \\
For such a large system it is more useful to consider a probability distribution $\rho(\vb{q},\vb{p})$ on the state-space. This is the point of view introduced by Gibbs. 
We will first consider the simplest, and by far the most common situation, where the Hamiltonian, $\mathcal {H}=\mathcal {H}(\vb{q},\vb{p})$ is the only observable. The maximum entropy distribution for this case is
\begin{align}
\rho (\vb{q},\vb{p}) = \frac{\rho_0(\vb{q},\vb{p})}{Z} \; \exp(-\frac{\mathcal {H}(\vb{q},\vb{p})}{kT}), \lbl{217.1} 
\end{align}
where the partition function is given by
\begin{align}
Z=Z(T)=\int_{\mathcal{R}^{6n}}d\vb{q}d\vb{p}\;\rho_0(\vb{q},\vb{p})\exp(-\frac{\mathcal {H}(\vb{q},\vb{p})}{kT}),\lbl{statmechZ1}
\end{align}
and where we have redefined the single Lagrange multiplier using
\begin{align}
\lambda = \inv{k \; T}. \lbl{218} 
\end{align}
In this formula, $k$ is the Boltzmann constant and $T$ is a new parameter which by definition is the thermodynamic temperature.
The parameter $T$ is determined by 
\begin{align}
 E&= \left<\mathcal{H}\right>, \nonumber \\
&\Updownarrow \nonumber \\ 
E&=k \; T^2 \; \prt{T} \ln Z ,\lbl{220}
\end{align}
where we have used the chain rule 
\begin{align}
\prt{\lambda} =- k \; T^2 \; \prt{T}, \lbl{221} 
\end{align}
in the general formula \rf{214}.

  Formula \rf{220} is in statistical mechanics and thermodynamics called the {\it equation of state}, and all thermodynamic statements that can be made about the system flows from this formula. The formula for equation of state may look innocent, in order to find it you merely need to take the derivative of the partition function, and partition function also looks innocent, after all it is just a function of one variable, the kind of function we study in first year calculus. However, in order to actually find an expression for this single variable function one needs to do the integral in formula \rf{statmechZ1}, and this is a multiple integral involving something like $10^{27}$ integration variables in typical situations! Clearly, an exact formula for the partition function can rarely be found. Approximate expressions where the large number of particles are used to ones advantage can more frequently be found, but pushing through calculations like these are as a rule extremely technical. More than one Nobel price has been handed out for developing feasible schemes for calculating the partition function. Given the level of complexity involved in calculating the partition function from the defining formula \rf{statmechZ1}, and the fact that the partition function simply is a function of one or a few variables,  it should come as no surprise that the most common approach to finding the equation of state is to fit parametrized functions to experimental data.

The maximum entropy distribution \rf{217.1} is recognized to essentially be the \ttx{Gibb's Canonical ensemble} from statistical physics.

The Gibb's ensemble is the foundation of statistical physics. All results in statistical physics flows from formula \rf{217.1}. Statistical physics is also the foundation of thermodynamics so all conclusions from that subject also flow from the Gibb's ensemble \rf{217.1}. In the thermodynamics context,  \rf{220} is, as we have already remarked, nothing but the \ttx{equation of state}. 

An interesting insight here is that the temperature of a thermodynamic system is in fact a Lagrange multiplier!! This is a profound insight that to this day has not been fully understood or explored.

From this example, it appears  useful to think of any application of the maximal entropy principle as an extension of the methods of statistical mechanics to systems that has absolutely nothing to do with the motion of mass points. 

This wide general applicability of the methods of statistical physics has  lead to deep questions and insights into the nature and significance of the assumption of equilibrium that appears to underline the application of the Gibb's ensemble in statistical physics. 

There is also the intriguing fact that the very same functional \rf{79} used in the maximum entropy principle,  is also the foundation of information theory which was discovered by Shannon in 1948. This connection between information theory and statistical mechanics (and thermodynamics) has lead to deep insights into the role of information in our fundamental physical theories. \\
As already discussed in the introduction, the general nature and wide applicability of the maximum entropy principle has been described well by E.T. Jaynes in many papers and the (unfinished) monumental book "Probability theory: The Logic of Science". 

\noindent As if all this is not impressive enough for one single principle, it is also a very intriguing fact that when one looks deep into the heart of fundamental physics, in the form of quantum field theory, one again finds an appropriately generalized form of  the partition function \rf{211}. The whole computational engine in the theory of quantum fields revolve around a generalized Gibb's ensemble! 

What on earth is going on...

\subsubsection{The problem of prior}

Note that formula \rf{217.1} does not uniquely define Gibbs ensemble because of the presence of the prior distribution $\rho_0$. The actual Gibbs ensemble corresponds to the choice $\rho_0=1$. When using the information theoretical approach to statistical mechanics and thermodynamics, like we do here, one should be very wary when it comes to the choice of the prior distribution. It is simply the most contentious  issue in the whole theory. One should ask pointed questions of justification for any proposed choice. What kind of information about the system is it based on, and is it the correct embodiment of said information? 

In fact, if one study expositions of statistical mechanics and thermodynamics which is based on the traditional objective dynamical approach to the subject,  one finds that the choice of what from the information theoretical point of view is the prior distribution, is much discussed. The reasons for choosing $\rho_0=1$ that have appeared through these discussions are, in our humble opinion, not very convincing.

The problem of determining the prior distribution has been at the center of probability theory
and statistics from the very start. The general rules of
probability theory tells us how to compute probabilities for
derived events from probabilities of primary events. The problem
of prior is concerned with the problem of assigning probabilities
to primary events. The assignment is supposed to reflect an
observers state of knowledge about the primary events. The
assignment should be the same for different observers with the
same state of knowledge but can be different for observers with
different states of knowledge \cite{Jaynes3}. In this sense
probability assignments are subjective
 \cite{jeffreys},\cite{cox1},\cite{cox2}. The problem of the prior
is how to turn states of knowledge into probability assignments.
The first solution to this problem was used by the very founders
of probability theory (Bernoulli and Laplace). If the observers
only knowledge of the primary events are their number, then a
uniform probability assignment should be used. This idea was later
named the principle of indifference by J. M. Keynes. Generalizing
this idea to countably infinite or even continuous spaces of
primary events has turned out to be very problematic. Laplace
himself used such a generalization is his work on probability
theory. His probability distribution was uniform and not
normalizable since it was defined on the whole real line. Using a
uniform distribution for representing indifference about a random
variable on a finite interval on the real line would seem to be
more reasonable, at least it is normalizable. However even in this
case serious problems arise as the well known Bertrand's paradox
shows. Problems and paradoxes arising from the various
generalizations of the principle of indifference to continuous
random variables played no small part in the creation
and for a long time complete dominance of the frequency interpretation\cite%
{Fisher} of probability theory.

The principle of maximum entropy appears first in the writings of
W. Gibbs  \cite{Gibbs} on thermodynamics and statistical physics
and later in the fundamental work on information theory by Shannon
\cite{Shannon}. However it was E. T. Jaynes \cite{Jaynes1} who
realized the real importance and general nature of the principle
of maximum entropy. In his hands it turned into a general method
for turning prior knowledge in the form of mean values for observables defined on finite state spaces, into prior probability assignments.

 Let us consider this simplest case in more detail. Let $\Omega =\{x_{1},x_{2},....,x_{n}\}$ be a finite
space of primary events. The algebra of possible events is the set of all subsets of $
\Omega$. A probability assignment on the set of primary events is a finite set of
numbers $p=\{p_{i}\}$ such that $0\leq p_{i}\leq 1$ and $\sum_{i=1}^{n}p_{i}=1$%
. Let $f_{1},...,f_{k}$ be real valued functions on $\Omega $. The
principle of maximum entropy states that if the means of the
functions $f_{1},..,f_{k}$ are known, $\left<f_{i}\right>=c_{i}$, one should,
among all probability assignments that satisfy the constraints,
pick the one that maximizes the entropy $S=-\sum_{i=1}^{n}p_{i}\ln
p_{i}$. The solution to this constrained maximization problem is, as we have seen, the maximum entropy distribution
\begin{align}
p=\frac{1}{Z(\lambda _{1},...,\lambda
_{k})}\exp\left(-\sum_{j=1}^{k}\lambda _{j}f_{j}\right),
\end{align}
where $Z$ is the partition function and is given by
\begin{align}
Z(\lambda _{1},...,\lambda _{k})=\sum_{i=1}^{n}\exp\left(-\sum_{j=1}^{k}\lambda _{j}f_{j}(i)\right).
\end{align}
Observe that for the particular situation where there are no constraints,
the principle gives $Z=n$ and the maximum entropy distribution is uniform
\begin{align}
p_{i}=\frac{1}{n}.
\end{align}
Thus, for an observer that only know that there are $n$ possible primary events, the maximum entropy distribution is exactly the one suggested by the principle of indifferent! The conclusion appeared to be that not only could the maximum entropy distribution tell us how to choose the best distribution in the presence of observed means of a finite number of observables, it could also tell us which distribution to choose when our ignorance is so profound that the only thing we know about a situation is the number of possible primary events. This distribution is of course exactly what we have called the prior distribution.  For
a time it looked as if the problem of prior was essentially
solved.  However continuous valued random variables again turned
out to be the Achilles heel. For finite spaces of events the
principle will give a unique probability assignment, but when
generalizing it to continuous random variables by taking a continuum limit of the finite discrete expression for the entropy, an unknown
probability density appears. The density appears because the continuum limit is not unique.
Different limiting expressions are found depending on how one approach the continuum through a countable set of discrete spaces.
The unknown probability distributions that appears essentially depends on how the discrete points bunch up in the limit. The meaning of this probability density became
clear when it was realized that it is the maximum entropy
distribution corresponding to no constraints. Thus it was
understood that in order to apply the principle of maximum entropy
one must start with a prior distribution. The principle of maximum
entropy could not determine the prior, it could only tell us how
to modify an already existing prior in order to satisfy
constraints in the form of mean values. It seemed as if one were
back to square one.

There {\it does} however exist  a systematic way to turn prior information on means of observables into prior
distributions, and it {\it does} involve the maximum entropy principle,  but not in the direct way just described.
In fact, after a certain reformulation  it will become evident that the problem of
 selecting a prior is not merely a side issue that has to be resolved in order to proceed with the real work of applying
  the maximum entropy principle, the problem of prior is the {\it only} issue as far as the maximum entropy principle is concerned.

In order to describe this reformulation of the principle of maximum entropy, we will return to the special case of statistical mechanics.
In the previous section we discussed the problem of specifying the prior in the context of statistical mechanics and expressed our doubt as to the justifications for making the standard choice $\rho_0(\vb{q},\vb{p})=1$. Even if we are doubtful about the justification for this particular choice,  it is clear that when we apply the maximum entropy principle  in statistical mechanics there is a physical context that certainly makes some choices of the prior more reasonable than othersr. By picking the Hamiltonian function as our observable we must also acknowledge that the system evolve according to the corresponding Hamiltonian equations. It is always the case that the the Hamiltonian function, $\mathcal{H}(\vb{q},\vb{p})$, which represents the energy, is a constant of the motion. Depending on the symmetries of the interaction,  Hamiltonian systems of equations may also have other conserved quantities. The generic situation is however that the energy is the only conserved quantity. We will assume that this is the case and let the corresponding Hamiltonian function $\mathcal{H}$ be our only observable. The maximum entropy distribution is now given by expression \rf{217} where $\rho_0$ is the prior distribution. It  is in the current context reasonable to impose the condition that the prior is a stationary solution to the corresponding Liouvillian equation. But this means that the prior distribution, $\rho_0=\rho_0(\vb{q},\vb{p})$,  is a conserved quantity for  the Hamiltonian system and since the Hamiltonian is the only independent conserved quantity for our generic Hamiltonian system we must have
\begin{align}
\rho_0(\vb{q},\vb{p})=f_0(\mathcal{H}(\vb{q},\vb{p})),\nonumber
\end{align}
where $f_0$ is an arbitrary function defined on the positive real line.
Using this fact we have from \rf{statmechZ1}
\begin{align}
Z(T)&=\int_{\mathcal{R}^{6n}}d\vb{q}\;d\vb{p}\rho_0(\vb{q},\vb{p})\exp(-\frac{\mathcal {H}(\vb{q},\vb{p}}{kT})\nonumber\\
&=\int_{\mathcal{R}^{6n}}d\vb{q}\;d\vb{p}f_0(\mathcal{H}(\vb{q},\vb{p}))\exp(-\frac{\mathcal {H}(\vb{q},\vb{p}}{kT})\nonumber\\
&=\int_0^{\infty} dE\;\exp(-\frac{E}{kT})\;f_0(E)\;\int_{\mathcal{H}=E}d\vb{q}\;d\vb{p}\nonumber\\
&=\int_0^{\infty} dE\;\exp(-\frac{E}{kT})\;\rho_0(E),\lbl{statmechZ2}
\end{align}
where we have defined 
\begin{align}
\rho_0(E)=f_0(E)\;\int_{\mathcal{H}=E}d\vb{q}\;d\vb{p}.\lbl{macroPrior}
\end{align}
The constraints on the {\it microscopic} prior distribution $\rho_0(\vb{q},\vb{p})$ has reduced our original maximum entropy principle on the extremely high dimensional space $\mathbb{R}^{6n}$ with the Hamiltonian as our observable,  to a maximum entropy problem on the real line where the coordinate on the line, $E$, is the observable and the {\it macroscopic} prior is given by \rf{macroPrior}. The maximum entropy distribution for this case is
\begin{align}
\rho(E)=\frac{\rho_0(E)}{Z(T)}\exp(-\frac{E}{kT}),\lbl{MaxEntOnLine}
\end{align}
where the partition function is given by
\begin{align}
Z(T)=\int_0^{\infty}dE\;\rho_0(E)\exp(-\frac{E}{kT}).\lbl{StatMechReducedPartitionFunction}
\end{align}
This simple situation where we apply the entropy principle to a low dimensional state space $\mathbb{R}^p$ and where the observables are the 
coordinate functions, $x_1,...,x_p$ on the space is not special at all, in fact this is the most common situation when we apply the maximum entropy principle and  other applications can almost always be reduced to this situation using an approach similar to the reduction from $\mathbb{R}^{6n}$ to $\mathbb{R}$ described for the case of statistical mechanics.

In most applications of probability theory in statistics there is
no underlying high dimensional space of primary events, $\Omega $,  like in statistical mechanics and other areas of physics,  and the random
variables are not some functions,  like the Hamiltonian, defined on this space. 

Thus in the typical case one can assume that $\Omega =\mathbb{R}^{p}$, where $p$ is a fairly small number,  and that the random variables are just the coordinate function on $\mathbb{R}^{p}$. 
The prior probability distribution is then a function, $\rho_0=\rho_0(x_1,...,x_p)$, on $\mathbb{R}^p$, and  the partition function is given by the formula
\begin{align}
Z(\lambda _{1},..,\lambda _{p})=\int_{\mathbb{R}^p}dx_1dx_2...dx_p\;\rho_0(x_1,...,x_p)\exp(-\sum_{j=1}^{p}\lambda _{j}x_{j}).\lbl{FromRho0ToZ}
\end{align}
The partition function is thus nothing but the multi dimensional Laplace
transform of the prior distribution. This relation can be inverted,
using analytical continuation and the multidimensional Fourier transform on the imaginary $\lambda_j$ axes, and thereby expressing the prior in terms of the partition function
\begin{align}
\rho _{0}(x_{1},..x_{p})=\frac{1}{(2\pi )^{p}}\int_{\mathbb{R}^{p}}d\lambda _{1}..d\lambda _{p}Z(i\lambda _{1},..,i\lambda _{p})\exp(i\sum_{j=1}^{p}\lambda
_{j}x_{j}).\lbl{FromZToRho0}
\end{align}The whole content of the maximum entropy principle is contained in the integral transforms \rf{FromRho0ToZ} and \rf{FromZToRho0} connecting the partition function and the prior distribution. This is the promised reformulation of the maximum entropy principle, and we understand now that the prior distribution is not merely a bit player in this drama, it is the {\it only} player. In the next section we will show how this reformulation of the maximum entropy principle gives us a method for solving the problem of prior, which we now see is the only remaining fundamental problem in the statistical modeling of natural or artificial systems. The material on the problem of prior in the next section has been previously published \cite{per} in a slightly different form.

\subsubsection{Solving the problem of prior using stochastic relations}

In probability theory and statistics, random variables are often grouped into
statistical quantities. These are certain algebraic
combinations of means of functions of the random variables. A
large set of such statistical quantities are in use, some simple
examples are
\begin{align}
&\left<x\right>\;\;\;\;\;\;\;\;\;\;\;\;\;\;\;\;\;\;\;\;\;\;\;\;\;\;\;\;\;\;\;\;\;\;\;\;\;\;\;\;\;\;\;\;\text{ The mean of x.}\\
&\left<x^{2}\right>-\left<x\right>^{2}\;\;\;\;\;\;\;\;\;\;\;\;\;\;\;\;\;\;\;\;\;\;\;\;\;\;\;\;\;\;\; \text{ The variance of x.}\\
&\left<x^{3}\right>-3\left<x\right>\left<x^{2}\right>+2\left<x\right>^{3}\;\;\;\;\;\;\;\;\;\;\;\text{The third cumulant.} \\
&\left<xy\right>-\left<x\right>\left<y\right>\;\;\;\;\;\;\;\;\;\;\;\;\;\;\;\;\;\;\;\;\;\;\;\;\;\;\;\;\;\text{The cross variance of x and y.}
\end{align}
All such quantities can systematically be expressed as functions of the form
$F(q_{1},..,q_{k})$ where the variables $q_{j}$ are means of monomials in
the random variables. We will define {\it stochastic relations} to be systems of
equations for the quantities $q_{j}$.%
\begin{align}
F_{i}(q_{1},..,q_{k}) =0\text{ \ \ \ \ }i=1,...,s.\lbl{StochasticRelation}
\end{align}
Such relations are common in probability and statistics. Examples
are zero mean, fixed variance, uncorrelated variables and
identities expressing higher order cumulants in terms of lower
ones. Identities such as the last ones in the previous list  are the fundamental tools used to construct theories of turbulence
in fluid, gases and elsewhere. They are also, in their quantum incarnations,the key tools used to find viable simplified models in solid state physics and material science.

  In the previous section we have seen that the maximum
entropy principle defines a Laplace transform that map the prior
distribution to a partition function. As a direct consequence of this
transformation we can express means of monomials in the random
variables in
terms of partial derivatives of the partition function. For example we have
\begin{align}
\left<f_i\right>&=-\frac{1}{Z}\partial_{\lambda_i}Z,\nonumber\\
\mathop{var}(f_i)&=\left<f_i^2\right>-\left<f_i\right>^2=-\frac{1}{Z^2}\partial_{\lambda}Z^2+\frac{1}{Z}\partial_{\lambda\lambda}Z.\nonumber
\end{align}

\noindent This means that the maximum entropy principle turns stochastic
relations into systems of partial differential equations for the
partition function and therefore imposes constraints on
the prior distribution.

    The problem is now how to describe the space of
solutions of these systems of partial differential equations. In
general, not all solutions to the equations can correspond to prior
probability distributions. From the definition of the partition function it is
for example clear that $Z(0)=1$ must hold for any acceptable solution.
 Finding necessary and sufficient conditions for functions to be the Laplace
transform of a probability distribution, and thus be acceptable solutions of
the systems of differential equations corresponding to stochastic relations,
 is not a simple matter, but
some results are known \cite {Rothaus}. We will not discuss this
problem but rather try to explicitly construct the solution space
or to say something useful about the structure of the solution space using
methods from the formal theory of differential equations .
Typically, the solution space is not a linear space and even when
it is, the dimension could easily be infinite. However, depending
on the number and types of stochastic relations the solution space
can end up being parametrized by a finite set of parameters or
even be a single point. In this last situation the stochastic
relations determine the prior uniquely. Note that in ordinary
(parametric) statistics finite parameter families of probability
distributions (Gaussian, Poissont, Bernoulli, t-distribution, etc)
are assumed to apply in given situations. From the point of view
discussed in these notes, this means that in ordinary statistics,
stochastic relations constrain the solution space enough for it to
be parameterized in terms of a finite number of parameters.
Nonparametric statistics correspond to the situation when the
solution space is so weakly constrained that it can not be
parameterized in terms of a finite number of parameters. Methods
from the theory of partial differential equations can in some
cases parameterize such weakly constrained solution spaces, not in
terms of real numbers, but in terms of arbitrary functions.
However for such weakly constrained solution spaces there is
another powerful tool available. This is the formal theory of
partial differential equations. The main object of study in this
theory is the infinitely prolonged hierarchy of a given systems of
differential equations. Thus one studies the infinite set of all
differential consequences of a given system of equations. Each
such differential consequence can be converted back into a
stochastic relation by using the relation between mean of
monomials and partial derivatives in reverse. One therefore gets
the corresponding hierarchy of stochastic relations that are
consequences of the original relations induced by the maximum
entropy principle and implemented through the Laplace transform.

In the remaining part of this subsection we will discuss several examples that
illustrate the method that has been outlined.

\paragraph{Stochastic relations for one random variable}
Essentially all families of distribution in use in parametric
statistics can be derived from simple stochastic relations
involving the mean, variance and skewness. In this section we show
some examples that support this statement.

\subparagraph{Delta distribution}
Let us consider the stochastic relation corresponding to a fixed
mean. It is
\begin{align}
\left<x\right>-q=0.
\end{align}
The Laplace transform convert this into the ordinary differential equation
\begin{align}
Z_{\lambda }=-qZ.
\end{align}
For this simple stochastic relation our system of partial differential
equations is a single linear ordinary differential equation. The solution
space is linear and parameterized by a single parameter
\begin{align}
Z(\lambda )=ae^{-q\lambda }.
\end{align}

\noindent The condition $Z(0)=1$ fixes the parameter $a$ to be one and we
have a unique solution. It is a simple matter to apply the inverse
transform \rf{FromZToRho0} to find the corresponding prior distribution
\begin{align}
\rho_0(x)& =\frac{1}{2\pi}\int_{-\infty}^{\infty}d\lambda Z(i\lambda)\exp(i\lambda x)\nonumber\\
&=\frac{1}{2\pi}\int_{-\infty}^{\infty}d\lambda\;\exp(-iq\lambda)\exp(i\lambda x)\nonumber\\
&=\frac{1}{2\pi}\int_{-\infty}^{\infty}d\lambda\;\exp(i(x-q)\lambda)\nonumber\\
&=\delta (x-q).
\end{align}

\subparagraph{Normal distribution}

The stochastic relation corresponding to constant variance  is
\begin{align}
\mathop{var}(x)=q,
\end{align}
and the corresponding differential equation is
\begin{align}
ZZ_{\lambda \lambda }-Z_{\lambda }^{2}-qZ^{2}=0.
\end{align}
This is a second order nonlinear ordinary differential equation.
The general solution of the nonlinear equation that satisfies the
requirement $Z(0)=1$ is
\begin{align}
Z(\lambda )=e^{-a\lambda +\frac{1}{2}q\lambda ^{2}}\text{, \ \ \ \
\ \ \ }a\in\mathbb{R}.
\end{align}
Using this partition function we can predict the mean of the random variable $x$ to be
\begin{align}
\left<x\right)&=-\frac{1}{Z(\lambda)}\frac{\partial Z(\lambda)}{\partial \lambda}\nonumber\\
&=a
\end{align}
and the corresponding prior distribution is found, using \rf{FromZToRho0}, to be

\begin{align}
\rho_0 (x)&=\frac{1}{2\pi }\int_{\mathbb{R}}d\lambda\;Z(i\lambda )\exp(i\lambda\;x)\nonumber\\
&=\frac{1}{2\pi}\int_{-\infty}^{\infty}d\lambda\;\exp(-ia\lambda  -\frac{1}{2}q\lambda ^{2})\exp(i\lambda\;x)\nonumber\\
&=\frac{1}{\sqrt{2\pi q}}e^{-\frac{(x-a)^{2}}{2q}}.
\end{align}
which is the normal distribution.

\subparagraph{Poisson distribution}

Let us consider the stochastic relation
\begin{align}
\mathop{var}(x)=<x>.
\end{align}

The corresponding differential equation is
\begin{align}
ZZ_{\lambda \lambda }-Z_{\lambda }^{2}+ZZ_{\lambda }=0.
\end{align}
This equation and most equations derived from stochastic relations simplify
considerably if we introduce a new function $\varphi $ through $Z=e^{\varphi
}$. The equation for $\varphi $ is
\begin{align}
\varphi _{\lambda \lambda }=-\varphi _{\lambda }.
\end{align}
This equation is easy to solve and the corresponding family of partition
functions satisfying, as always, the constraint $Z(0)=1$ is
\begin{align}
Z(\lambda )=e^{a(e^{-\lambda }-1)}.
\end{align}
The corresponding prior distribution is found using \rf{FromZToRho0} to be supported on $\Omega =\{0,1,2,....\}$ and
is of the form
\begin{align}
\rho_0 (k)=\frac{e^{-a}a^{k}}{k!}.
\end{align}
This is the Poisson distribution.

\subparagraph{Gamma distribution}

Let us consider a stochastic relation
\begin{align}
\mathop{var}(x)=\frac{1}{k}{\left<x\right>}^{2}\text{ \ \ }k>0.
\end{align}
Expressed in terms of $\varphi $ the corresponding differential equation is
\begin{align}
\varphi _{\lambda \lambda }=\frac{1}{k}\varphi _{\lambda }^{2}.
\end{align}
The general solution of this equation gives the following family of
partition functions
\begin{align}
Z(\lambda )=(1-a\lambda )^{-k}\text{ \ \ }a>0.
\end{align}
The correspondingprior  distribution is supported on $\Omega =(0,\infty )$ and is given by 
\begin{align}
\rho_0 (x)=x^{k-1}\frac{e^{-\frac{x}{a}}}{a^{k}\Gamma (k)}.
\end{align}
This is the Gamma distribution

\subparagraph{Bernoulli and Binomial distribution}

Let the variance be the following quadratic function of the mean
\begin{align}
\mathop{var}(x)=\left<x\right>(1-\left<x\right>).
\end{align}
The corresponding differential equation for $\varphi $ is
\begin{align}
\varphi _{\lambda \lambda }=-\varphi _{\lambda }(1+\varphi _{\lambda }).
\end{align}
The solution of the equation gives the \ following family of partition
functions
\begin{align}
Z(\lambda )=p+qe^{-\lambda }\text{ \ \ \ \ \ }p+q=1.
\end{align}
The corresponding distribution is supported on $\Omega =\{0,1\}$
and is given by $\rho (0)=p$, $\rho (1)=q$. This is the Bernoulli
distribution. If we generalize the stochastic relation to
\begin{align}
\mathop{var}(x)=\left<x\right>(1-\frac{1}{n}\left<x\right>).
\end{align}
where $n$ is a natural number we get the differential equation
\begin{align}
\varphi _{\lambda \lambda }=-\varphi _{\lambda }(1+\frac{1}{n}\varphi
_{\lambda }).
\end{align}
The partition function is found to be
\begin{align}
Z(\lambda )=(p+qe^{-\lambda }\text{ )}^{n}\text{\ \ \ \ \ }p+q=1.
\end{align}
The corresponding prior distribution is now found to be supported on $\Omega =\{0,1,...n\}$ and is
on this domain given by
\begin{align}
\rho_0 (k)=\binom{n}{k}p^{k}q^{n-k}.
\end{align}
This is the Binomial distribution.

\bigskip

\paragraph{Stochastic relations for more than one random variable}

When the number of random variables become larger than one,
stochastic relations in general leads to systems of nonlinear
partial differential equations. Unless the number and type of
relations is right, it is impossible to describe the solution
space in terms of a finite number of parameters. This lead us into
the domain of nonparametric statistics. This is the domain where
the methods from the formal theory of differential equations comes
into play. It is not possible to give nontrivial applications of
the theory here and  we will therefore limit ourselves to
two simple examples.

\subparagraph{The Multinomial distribution}

Let $x_{1},...x_{n}$ be $n$ random variables and consider the
following
system of stochastic relations
\begin{align}
\mathop{var}(x_{i}) &=\left<x_{i}\right>(1-\frac{1}{n}\left<x_{i}\right>)\text{ \ }
i=1,..n\;, \nonumber\\
\mathop{cov}(x_{i},x_{j}) &=-\frac{1}{n}\left<x_{i}\right>\left<x_{j}\right>\text{ \ \ }
i,j=1,...n,\text{ \ \thinspace }i\neq j.
\end{align}
The corresponding system of partial differential equations is%
\begin{align}
\varphi _{\lambda _{i}\lambda _{i}} &=-\varphi _{\lambda _{i}}(1+\frac{1}{n}
\varphi _{\lambda _{i}}), \nonumber\\
\varphi _{\lambda _{i}\lambda _{j}} &=-\frac{1}{n}\varphi _{\lambda
_{i}}\varphi _{\lambda _{j}}.
\end{align}
The second part of the  system of equations has general solutions of the
form $\varphi =n\ln (\theta )$ where $\theta (\lambda _{1},..,\lambda
_{n})=\sum_{i=1}^{n}\theta _{i}(\lambda _{i})$. \ Inserted into the first
part of the system this form of $\varphi $ easily gives the partition
function corresponding to the multinomial distribution. This system of
relations thus constrained the space of solutions so much that it could be
describes in terms of a finite number of parameters.

\subparagraph{Stochastic relations for the mean}

For a single random variable, stochastic relations involving only
the mean gives distributions located on a finite number of points.
For more than one random variable such relations gives rise to
nonparametric statistics, or solution spaces parameterized by
functions. The theory of partial differential equations can be
used to give a full description of these solution spaces. As an
example of such a relation consider the case of two
random variables whose means are constrained to be on a circle of radius $r$.
\begin{align}
{\left<x\right>}^{2}+{\left<y\right>}^{2}=r^{2}.
\end{align}
The corresponding partial differential equation is i terms of $\varphi $
\begin{align}
\varphi _{\lambda }^{2}+\varphi _{\mu }^{2}=r^{2},
\end{align}
and is known from optics as the {\it Eiconal} equation.The following $Z$ is in the solution space
\begin{align}
Z=e^{r\sqrt{\lambda ^{2}+\mu ^{2}}}.
\end{align}
This partition functions predicts that the following stochastic relation
should hold%
\begin{align}
\mathop{var}(x)=\left( \frac{\left<y\right>}{\left<x\right>}\right) ^{2}\mathop{var}(y).
\end{align}
The partial differential equation has, however, infinitely many solutions.
The method of characteristics can be used to describe the complete solution
space. In order to derive stochastic relations that holds for all $Z$ in the
solution space, these are the ones that can be said to be consequences of
the of the circle constrain, we should consider differential prolongations
of the original differential equation. The first prolongation is the system
\begin{align}
\varphi _{\lambda }^{2}+\varphi _{\mu }^{2} &=r^{2}, \\
\varphi _{\lambda }\varphi _{\lambda \lambda }+\varphi _{\mu }\varphi _{\mu
\lambda } &=0, \\
\varphi _{\lambda }\varphi _{\lambda \mu }+\varphi _{\mu }\varphi _{\mu \mu
} &=0,
\end{align}
and this system implies that%
\begin{align}
\varphi _{\lambda \lambda }=\left( \frac{\varphi _{\mu }}{\varphi _{\lambda }%
}\right) ^{2}\varphi _{\mu \mu }.
\end{align}

Translated into stochastic relations this is exactly the one we derived for
the special solution $\varphi =r\sqrt{\lambda ^{2}+\mu ^{2}}$ and it thus
holds for all solutions. It is of considerable interest to find a finite set
of basic stochastic relations that through some construction procedure
implies all consequences of some given system of stochastic relations. This
is exactly the kind of question addressed in the formal theory of partial
differential equations and the tools developed there can now through the
maximum entropy principle be brought into the area of nonparametric
statistics.

\subsubsection{Thermodynamic pressure and its cousins}\lbl{pressure}
We will now investigate an important consequence of the fundamental variational identity \rf{FundamentalVariationalIdentity} for the thermodynamic case when the total energy is the only observable. For this special case the variational identity turns into
 \begin{align}
\left< \delta_\Gamma\mathcal{ H} \right>&= -kT\;\delta_\Gamma \ln Z.\lbl{FundamentalVariationalIdentity1}
  \end{align}
  The force acting at a boundary point $\vb{x}$ can, taking into account the fact that the state of the system is determined by the position and momenta of all the $n$ particles comprising the system, be given by a function
  \begin{align}
  \vb{F}&=\vb{F}(\vb{p},\vb{q},\vb{x}),\;\;\;\;\;\vb{x}\in\Gamma,\;(\vb{p},\vb{q})\in\mathbb{R}^{6n}.\lbl{BoundaryForce}
  \end{align}
  A small deformation of the boundary is determined by an infinitesimal deformation vector  field $d\vb{r}_{\Gamma}$ defined on the boundary $\Gamma$. The change in total energy induced by this deformation is given by
  \begin{align}
\delta_{\Gamma}\mathcal{H}(\vb{p},\vb{q};\Gamma) &=-\int_{\Gamma}\;d\vb{x}\;\vb{F}(\vb{p},\vb{q},\vb{x})\cdot d\vb{r}_{\Gamma}.
  \end{align}
  The fundamental variational identity \rf{FundamentalVariationalIdentity} now gives
 \begin{align}
\int_{\Gamma}\;d\vb{x}\;\left<\vb{F}(\vb{p},\vb{q},\vb{x}) \right>\cdot d\vb{r}_{\Gamma}&= kT\;\delta_\Gamma \ln Z.\lbl{FundamentalVariationalIdentity2}
  \end{align}  
  We will in the following only consider the common situation defined by
  \begin{align}
  \left<\vb{F}(\vb{p},\vb{q},\vb{x}) \right>&=p(\vb{x})\;\vb{n},\lbl{PressureAssumption}
  \end{align}
 where $\vb{n}$ is the unit normal for the surface $\Gamma$.  Note that by definition, $p$ is now the pressure for the system. For this case \rf{FundamentalVariationalIdentity2} turns into the identity
 \begin{align}
\int_{\Gamma}\;d\vb{x}\;p(\vb{x})\;\vb{n}\cdot d\vb{r}_{\Gamma}&= kT\;\delta_\Gamma\ln Z.\lbl{FundamentalVariationalIdentity2}
\end{align}
Let us first consider the case of a smooth surface, and for this kind of surface,  let us consider an infinitesimal variation of the surface that is a pure expansion or contraction. This means that $d\vb{r}_{\Gamma}=\vb{n}ds$. For this kind of variation the fundamental variational identity \rf{FundamentalVariationalIdentity2} takes the form
 \begin{align}
ds\;\int_{\Gamma}\;d\vb{x}\;p(\vb{x})&= kT\;\delta_\Gamma\ln Z.\lbl{FundamentalVariationalIdentity3}
\end{align}
Using the fact that the volume spanned by the deformation is $d_{\Gamma}V=A(\Gamma)ds$, where $A(\Gamma)$ is the area of the surface, we have 
\begin{align}
\left<p\right>_{\Gamma}= kT\;\frac{\delta_\Gamma\ln Z}{d_{\Gamma}V},\lbl{FundamentalVariationalIdentity3.5}
\end{align}
where $\left<p\right>_{\Gamma}$ is the average of the pressure over the surface of the cavity.

For some important cases the partition function depends on the surface only through the volume. For this situation we have
\begin{align}
\delta_\Gamma\ln Z=\frac{\partial\ln Z}{\partial\;V}d_{\Gamma}V,\nonumber
\end{align}
so that 
 \begin{align}
\left<p\right>_{\Gamma}= kT\;\frac{\partial\ln Z}{\partial\;V}.\lbl{FundamentalVariationalIdentity4}
\end{align}
This is the standard formula for the thermodynamic pressure that one finds in any textbook. It is very frequently true that, independently of the shape, the partition function for large cavities depends only on the volume of the cavity. This may however not be the case for smaller cavities and for such cases we must retreat to the more general identity \rf{FundamentalVariationalIdentity3.5}. It is easy to verify that the indentity \rf{FundamentalVariationalIdentity4} holds for any surface smooth or not. If the deformation is a pure expansion or contraction of a part of the surface defined by $\Gamma_0\subset\Gamma$, we also get the identity \rf{FundamentalVariationalIdentity4}, but now with $\Gamma\rightarrow\Gamma_0$.

The general variational identity for the entropy \rf{dS5} takes for the particular case discussed in this section the form
\begin{align}
\delta S&=\frac{1}{kT}d\left<\mathcal{H}\right>-\frac{1}{kT}\left< \delta_\Gamma\mathcal{ H} \right>\nonumber\\
&=\frac{1}{kT}d\left<\mathcal{H}\right>+\delta_\Gamma \ln Z\nonumber\\
&=\frac{1}{kT}d\left<\mathcal{H}\right>+\frac{1}{kT}\left<p\right>_{\Gamma}d_{\Gamma}V,\nonumber
\end{align}
which can be rewritten as
\begin{align}
kTdS=dE+\left<p\right>_{\Gamma}d_{\Gamma}V,\lbl{dS6}
\end{align}
where we have used \rf{FundamentalVariationalIdentity3.5} and where now $E=\left<\mathcal{H}\right>$ is the energy of the system. We recognize \rf{dS6} as one of the fundamental formulas from conventional thermodynamics.

In this section we have done the derivation of the formulas for the Thermodynamical pressure for the case of a classical system. However, the derivation of the pressure formula for the case of quantum systems leads the exact same formulas. If there are more observables in addition to the energy, for example total momentum and/or total angular momentum, the pressure formulas must be generalized. The derivation of the generalizations follow the pattern laid down in this section.

\subsection{The maximum entropy principle for quantum systems}
We have in the first section of these notes introduced the thermodynamical formalism in the context of classical physics and classical observables. It involved a state space that was finite or at least finite dimensional, and the challenge was to determine which probability distribution on the state space is the best to use, given the means of a finite number of observables of the system. The solution to this problem was to choose the probability distribution that maximized the entropy functional \rf{78} under the constraints determined by the given means. At the face of it, for a quantum system, the situation appears to be very different. For this case the state space is an infinite dimensional Hilbert space and the full information that an observer has is encoded in the \ttx{density operator} for the system. This is a self-adjoint positive operator $\hat\rho$, on the Hilbert space with trace equal to one.
\begin{align}
\mathrm{Tr}(\hat\rho)=1.\nonumber
\end{align}
The expectation value of any quantum observable, $\hat A$ is by definition
\begin{align}
\left<\hat A\right>=Tr(\hat\rho\hat A).\lbl{QuantumMean}
\end{align} 
The question one poses is which density operator should be used if we only know the expectation value of a finite number of quantum observables $\hat A_i$.
\begin{align}
\left<\hat A_i\right>&=a_i,\;\;\;i=1,2,...p.\lbl{Constraints}
\end{align}
Even though there are real differences between the classical and the quantum case, much is also the same. 

The analog of the Gibbs entropy measure \rf{79} is the Von Neuman entropy measure for density operators given by
\begin{align}
S(\hat\rho)=-Tr(\hat\rho\ln\hat\rho).\lbl{NeumanEntropy}
\end{align}
 The solution to the question posed on the previous page proposed by the Maximum entropy method is to choose the density operator that maximize the Von Neuman entropy while satisfying the constraints \rf{Constraints}. It is a simple exercise to show that the unique solution to this maximization problem is the following density operator
 \begin{align}
 \hat\rho=\frac{1}{Z(\lambda_1,...,\lambda_p)}\exp(-\sum_i\;\lambda_i\hat A_i).\lbl{MaxEntDensityOperator}
 \end{align}
 This operator is the \ttx{maximum entropy density operator}. The function $Z$ is the partition function and is given by
 \begin{align}
 Z(\lambda_1,...,\lambda_p)=Tr\left\{\exp(-\sum_i\;\lambda_i\hat A_i)\right\}.\lbl{PartitionFunction}
 \end{align}
 Arguing like in the classical case we find that much of the thermodynamic formalism is the same as before. Spesifically we have
 \begin{align}
 S&=\ln Z+\sum_{i=1}^p\;\lambda_i\left<\hat A_i\right>,\nonumber\\
 \left<\hat A_i\right>&=-\frac{\partial \ln Z}{\partial \lambda_i},\nonumber\\
 \lambda_i&=\frac{\partial S}{\partial \left<\hat A_i\right>},\\
\sum_i\lambda_i \left<\frac{\partial \hat A_i}{\partial \alpha}\right>&=-\frac{\partial \ln Z}{\partial \alpha},\lbl{QThermodynamicFormalism}
\end{align}
where in the last identity we assume that all the observables depend on some parameter $\alpha$. In our application of this formalism the operators will depend on the surface enclosing a domain $D$, and for this case we get a quantum analog to the classical formula \rf{FundamentalVariationalIdentity}
 \begin{align}
 \delta_\Gamma\ln Z=-\mathlarger{\sum}_j \; \lambda_j \;\left< \delta_\Gamma \hat A_j\right>.\lbl{QuantumFundamentalVariationalIdentity}
  \end{align}
  As we can see, much of the thermodynamic formalism is the same for the classical and the quantum case. However some things are different, or they at least appear to be different. In the classical case we can find correlations between different observables by computing mixed partial derivatives of the partition function as shown in \rf{215} and \rf{216}. In the quantum case this is problematic  unless the operators commute. For the case of two observables $\hat A$ and $\hat B$ we have for example
  \begin{align}
  \frac{\partial^2 Z}{\partial_{\mu}\partial_{\lambda}}= \frac{\partial^2 Z}{\partial_{\lambda}\partial_{\mu}}=2\left<\hat A\hat B+\hat B\hat A\right>-\left<\hat A\right>\left<\hat B\right>.\nonumber
  \end{align}
 In a sense this should not come as a surprise. The reason for this is that $\hat A\hat B$ is not in general Hermitian even if both $\hat A$ and $\hat B$ are. Thus $\left<\hat A\hat B\right>$ is not something that you could ever measure, so it does not make sense to try to predict it. However, $\hat A\hat B+\hat B\hat A$ $\text{\it is}$ a Hermitian operator and thus $\left<\hat A\hat B+\hat B\hat A\right>$ is something one would want to predict. And,  this is exactly what you would be able to predict using the thermodynamical formalism.
  
  There is another important way in which the classical and quantum cases are different; namely the question of how to include prior information about the system into the thermodynamical formalism. In the classical case this was taken care of by using the entropy measure \rf{79} that included the prior distribution $\rho_0$. We have seen how maximization the entropy in the context of statistical mechanics leads to the distribution 
\begin{align}
\rho (\vx_1,...,\vx_n, \vb{p}_1,..., \vb{p}_n) = \frac{\rho_0}{Z} \; \exp(-\frac{\mathcal{H}}{kT}), \lbl{217} 
\end{align}
which we recognized to be the Gibbs canonical ensemble. This is however not entirely correct, the canonical ensemble corresponds to the case when we have a uniform prior. We could have gotten this distribution directly from a maximization of the functional
\begin{align}
S(\rho) = - \int_{\mathbf{R}^n} dV \; \rho \; \ln (\rho). \lbl{GibbsEntropy} 
\end{align}
This is in fact the functional used by Gibbs in his foundational studies of thermodynamics. It is this functional that is called the Gibbs entropy measure. The functional we introduced in \rf{79} measure by definition the relative entropy of $\rho$ with respect to $\rho_0$. It is also denoted by other names in the research literature.

  The Von Neuman entropy introduced in \rf{NeumanEntropy} is the direct analog of the Gibbs entropy measure \rf{GibbsEntropy}. However, in contrast to the classical case, there does not exists a measure of information for quantum systems that is universally recognized to be the best measure to use in all cases where there is  prior information to take into account. Many different kind of measures has been studied by physicists and mathematicians over the years. Today these questions are intensely pursued in the topical field of quantum information theory. 

Note that the classical relative entropy measure can be written in the form
\begin{align}
S(\rho|\;\rho_0) = - \int_{\mathbf{R}^n} dV \; \rho \; (\ln (\rho-\ln\rho_0)). \lbl{RelativeEntropy} 
\end{align}
One approach to a quantum version of relative entropy is to directly generalize \rf{RelativeEntropy} to the quantum case. This gives us for any pair of density operators $\hat\rho$ and $\hat\rho_0$ the relative quantum entropy in the form
\begin{align}
S_Q(\hat \rho|\;\hat\rho_0) = - Tr( \hat \rho \; (\ln \hat \rho-\ln\hat\rho_0)). \lbl{QRealtiveEntropy} 
\end{align}
One could now guess that the corresponding maximal entropy distribution, when the only observable is the total energy, will take the form
\begin{align}
\hat\rho = \frac{\hat\rho_0}{Z} \; \exp(-\frac{\mathcal{\hat H}}{kT}), \lbl{QMaxEntDensityOperator} 
\end{align}
However, a formula like this can not possibly  be correct because the right hand side of \rf{QMaxEntDensityOperator} is not a Hermitian operator unless the prior $\hat\rho_0$ commutes with the total energy operator $\mathcal{\hat H}$.

  Here we will not pursue these matters, but rather use the Neuman entropy measure as a basis for the theory. But we should keep in mind that there is a real issue here concerning the general validity of the thermodynamical formalism for quantum systems in cases when there is prior information present.

\section{Canonical quantization of electromagnetic fields in a cavity}
Quantum theory, as we know it today, originated from the investigation of electromagnetic fields confined to a cavity. At first however, the insights gained from the study of this system by Planck, Einstein and others did not lead  to a quantum description of electromagnetic fields in cavities, but rather to the quantum description of atoms and molecules by Schr\o dinger and Heisenberg. This is the theory that today is known as  {\it Quantum Mechanics}.  Compared to the quantum description of fields(QFT), the electromagnetic field in particular, the quantum theory of atoms and molecules is conceptually and technically much simpler. Of course, calculating quantum properties of atoms and molecules is not in any way simple, it is just that QFT is so very hard. 

By the end of the 1920s, quantum mechanics was well established and widely regarded as the correct theory of atoms and molecules. The general principles and abstract mathematical structures underlying quantum mechanics was at this time known, if not entirely understood. Books published by P. A. M. Dirac in 1930\cite{Dirac2} and John von Neuman in 1932\cite{VonNeuman} gave a precise description of these underlying structures and principles. What they describe in these two books is what today is known as {\it Quantum Theory}. It has been realized over time that this theory, Quantum Theory, is not actually a physical theory at all, but rather a general mathematical scheme that  {\it any} fundamental physical theory must  conform to.

  W. Heisenberg, Max Born and Pascual Jordan were in 1926 the first physicists to construct a viable quantum theory of the electromagnetic field\cite{Born}. The theory they constructed is what we  today call a quantum theory of {\it free fields}. 
  
    Their approach was to expand the electromagnetic field in modes in such a way that the total  energy of the system, which in classical mechanics is known as the {\it Hamiltonian} for the system, became a sum of an infinite set of independent harmonic oscillator Hamiltonians. Each of these harmonic oscillators could then be quantized using the Heisenberg canonical quantization rules for bosonic particle systems that that were well understood at the time.
    
      Of course, for a quantum theory of the electromagnetic field to become a useful and flexible tool for the description of nature, the interaction between electrons and the electromagnetic fields had to be taken into account and given a description that conformed to the rules of Quantum Theory. This was achieved by Dirac in 1927\cite{Dirac1}. 

In these notes we will follow the simple approach introduced by Heisenberg, Born and Jordan in 1926, and we will start by considering the very simplest case of an empty cavity with perfectly reflecting walls. 
Maxwell's equations for the electromagnetic field in an empty cavity, $D$ are
\begin{align}
\curl \bf{E}+\mu_0\partial_t\vb{H}&=0,\nonumber\\
\curl\vb{H}-\epsilon_0\partial_t\vb{E}&=0,\nonumber\\
\div\vb{E}&=0,\nonumber\\
\div\vb{H}&=0.\lbl{Maxwell1}
\end{align}
These equations are supplemented with boundary conditions on the perfectly reflecting bounding surface $\Gamma$ defining the cavity $D$
\begin{align}
\vb{n}\times\vb{E}(\vb{x},t)&=0,\nonumber\\
\vb{n}\cdot\vb{H}&=0,\;\;\;\;\forall \bf{x}\in\Gamma.\lbl{BC1}
\end{align}
From the Maxwell equations \rf{Maxwell1} it is easy to show that the the electric field solves the equation
\begin{align}
\curl(\curl\vb{E})&=(\frac{\omega}{c})^2\vb{E},\nonumber\\
\div\vb{E}=0.\lbl{AProblem}
\end{align}
The magnetic field solves the exact same equation. 
 In order to solve this problem we introduce an eigenvalue problem
 \begin{align}
\curl(\curl\vb{u})&=(\frac{\omega}{c})^2\vb{u},\nonumber\\
\div\vb{u}=0,\nonumber\\
\vb{n}\times\vb{u}=0.\lbl{EigenvalueProblem}
\end{align}
 One can show\cite{Jones} that this eigenvalue problem has a complete set of orthogonal eigenfunction $\vb{u}_{\vb{m}}$  and corresponding eigenvalues $(\frac{\omega_{\vb{m}}}{c})^2$. We will assume that the eigenfunctions has been normalized so that
 \begin{align}
 \int_Dd\vb{x}\;\vb{u}_{\vb{m}}\cdot\vb{u}_{\vb{m}'}=\delta_{\vb{m}\;\vb{m}'}.\nonumber
 \end{align}
 
\noindent  Observe however, that there appears to be some kind of problem here. For a particular eigenfunction we have a corresponding electric field mode
 \begin{align}
 \vb{E}_{\vb{m}}(\vb{x},t)&=\vb{u}_{\vb{m}}(\vb{x})e^{-i\;\omega_{\vb{m}}\; t},\nonumber
\end{align}
 and the boundary conditions on the eigenfunctions $u_{\vb{m}}$ ensure that the field mode satisfy the boundary condition $\vb{n}\times\vb{E}_{\vb{m}}=0$. However it is \ttx{not} obvious that the corresponding magnetic field satisfy the boundary condition $\vb{n}\cdot\vb{H}=0$. The reason for this is that it is not in general true that $\vb{n}\cdot\curl\vb{f}=0$ is implied by $\vb{n}\times\vb{f}=0$ for a vector field $\vb{f}$ defined on a domain $D$ bounded by a surface whose unit normal is $\vb{n}$.
 The resolution of this apparent problem is somewhat subtle. The fact of the matter is that the result derived \cite{Jones} is a pure \ttx{existence} result, it is not  constructive. It only ensures the existence of a set of electric field modes
 \begin{align}
  \vb{E}_{\vb{m}}(\vb{x},t)&=\vb{u}_{\vb{m}}(\vb{x})e^{-i\;\omega_{\vb{m}}\; t},\nonumber
 \end{align}
  such that any solution to the problem \rf{Maxwell1},\rf{BC1} can be written as an expansion in $\vb{E}_{\vb{m}}$ and $\curl\vb{E}_{\vb{m}}$. In order to actually \ttx{construct} the modes in any particular case we will end up with an situation where one of the two boundary conditions \rf{BC1} is automatically satisfied and where imposing the second one will determine the eigenvalues.
  
  Using the modes $\vb{u}_{\vb{m}}$ we can write down following expansions for the electric and magnetic field
 \begin{align}
  \vb{E}(\vb{x},t)&=-\sum_{\vb{m}}\frac{\dot q_{\vb{m}}(t)}{\sqrt{\epsilon_0}}\;\vb{u}_{\vb{m}}(\vb{x}),\lbl{EExpansion1}\\
  \vb{H}(\vb{x},t)&=\sum_{\vb{m}}\frac{q_{\vb{m}}(t)}{\mu_0\sqrt{\epsilon_0}}\;\curl \vb{u}_{\vb{m}}(\vb{x}),\lbl{HExpansion1}
  \end{align}
  where
  \begin{align}
  q_{\vb{m}}''(t)=-\omega_{\vb{m}}q_{\vb{m}}(t).\lbl{HarmonicOscillator1}
  \end{align}
  This is the equation for a harmonic oscillator, and it, together with \rf{EigenvalueProblem} ensure that the expansions \rf{EExpansion1},\rf{HExpansion1} for the electric and magnetic fields  represents the general solution of the Maxwell equations in a cavity with perfectly conducting walls.
  
  \noindent The total electromagnetic energy in the cavity is 
  \begin{align}
\mathcal{H}= \frac{1}{2}\;\int_Dd\vb{x}\left(\epsilon_0\vb{E}\cdot\vb{E}+\mu_0\vb{H}\cdot\vb{H}\right).\lbl{TotalEnergy}
  \end{align}
  \noindent Inserting the expansions for the electric and magnetic fields, and using the orthonormality of the modes and their defining equation \rf{EigenvalueProblem}, we immediately get
  \begin{align}
  \mathcal{H}=\sum_{\vb{m}}\frac{1}{2}\left(\dot q_{\vb{m}}^2+\omega_{\vb{m}}^2q_{\vb{m}}^2\right)\lbl{OscillatorExpansion1}
  \end{align}
  \noindent We observe that each term in this Hamiltonian is the Hamiltonian for a harmonic oscillator of unit mass. We have now realized a description of the electromagnetic field as an infinite system of uncoupled harmonic oscillators. This is the starting point for the Heisenberg, Born, Jordan approach to quantizing free fields in a cavity.
  
   However,  before we proceed with the quantization of the harmonic oscillators, we will generalize the situation in a way that will cover all cases of interest to us in these notes.

    As a matter of fact, the Heisenberg, Born, Jordan approach to quantizing the electromagnetic field does not only work for an empty cavity, it also works for a cavity containing electrons and atoms. The restriction is that there are no free electrons; all electrons are bound to atoms. Furthermore, we require that the interaction between the electromagnetic field and and the atoms can be accurately modeled by a macroscopic {\it real} refractive index, thus no absorption is allowed. A further essential restriction is that the material response is {\it  local in time}; thus no material dispersion is allowed. We will however allow for the possibility that the refractive index vary from point to point in the cavity. It is not an essential requirement, but we will in these notes also assume that the material is nonmagnetic. 
    
    The assumptions made amounts the following relations connecting the fields $\vb{B}$ and $\vb{D}$ to $\vb{H}$ and $\vb{E}$
 \begin{align}
 \vb{B}&=\mu_0\vb{H},\nonumber\\
 \vb{D}&=\epsilon_0n(\vb{x})^2\vb{E}.\lbl{DERealtions}
 \end{align}
 We will be considering a situation where one or several separate objects with different dielectric constants are contained in a cavity which has  some constant background index. Let the different objects occupy domains $D_1,...D_p$ in the cavity with refractive indices $n_1,...,n_p$. The boundary of object $j$ is denoted by $\Gamma_j$. Let $D_0$ be the domain inside the cavity that remains after the domains $D_1,...D_p$ has been subtracted from the domain defining the whole cavity, which is denoted by $D$. The background refractive index in $D_0$ is denoted by $n_0$. The outer boundary of the cavity is denoted by $\Gamma_0$. By common convention the unit normal for $\Gamma_0,\Gamma_1,...\Gamma_p$ are pointing out of the domains they define. The geometry of the situation is summarized in the following formulas
 \begin{align}
D&=\cup_{j=0}^p D_j,\nonumber\\
\partial D_j&=\Gamma_j,\;\;\;j=1,...p,\nonumber\\
\partial D_0&=\Gamma_0-\cup_{j=1}^p\Gamma_j,\lbl{Geometry}
\end{align}
where the minus sign signify that $\Gamma_j$, as part of the boundary of $D_0$, has a unit normal that points into the domain $D_j$ since the inside of $D_j$ is part of the outside of $D_0$.

  It is easy to show that the electric field satisfy the equation 
\begin{align}
\curl(\curl\vb{E})&=(\frac{\omega}{c})^2n_j^2\vb{E}\;\;\;\;\;\vb{x}\in D_j,\nonumber\\
\div\vb{E}=0,\lbl{EProblem}
\end{align}
as does the magnetic field. With this equation in mind we introduce the following eigenvalue problem  
\begin{align}\left. \begin{array}{ccc}
&\curl(\curl\vb{u})=(\frac{\omega}{c})^2n_j^2\vb{u}\\
                   &                      \\
&\div\vb{u}=0
\end{array} \right\} \vb{x}\in D_j\;\;\;\;j=1,...p.\lbl{EigenValueProblem2}
\end{align}

\noindent In order to complete the description of the eigenvalue problem we must supply boundary conditions on all bounding surfaces $\Gamma_0,\Gamma_1,...,\Gamma_p$. 

For the boundaries internal to the cavity $\Gamma_1,...,\Gamma_p$ we will assume the usual electromagnetic interface conditions for dielectric materials. For the external boundary $\Gamma_0$ we will either assume that the boundary is perfectly reflecting and thus use the same boundary conditions as in \rf{EigenvalueProblem}, for this case the bounding surface can take on any shape, or for the special case of a rectangular cavity we will use a combination of periodic boundary conditions connecting fields on all or some pairs of opposing surfaces, and use perfectly reflecting surface conditions for the remaining faces. The eigenvalue problem is now fully specified and we will assume that this eigenvalue problem has a complete set of orthonormal eigenfunctions $u_{\vb{m}}$, where the  orthonormalization condition is defined by
\begin{align}
&\int_{D}d\vb{x\;}n(\vb{x})^2\vb{u}_{\vb{m}}\cdot\vb{u}_{\vb{m'}}\nonumber\\
&=\sum_{j=0}^p\int_{D_j}d\vb{x}\;n_j^2\vb{u}_{\vb{m}}\cdot\vb{u}_{\vb{m'}}=\delta_{\vb{m}\vb{m'}}.\lbl{Normalization}
\end{align}
It is simple to show that the general solution of the Maxwell equations for this more complicated cavity situation still are of the form
 \begin{align}
  \vb{E}(\vb{x},t)&=-\sum_{\vb{m}}\frac{\dot q_{\vb{m}}(t)}{\sqrt{\epsilon_0}}\;\vb{u}_{\vb{m}}(\vb{x}),\nonumber\\
  \vb{H}(\vb{x},t)&=\sum_{\vb{m}}\frac{q_{\vb{m}}(t)}{\mu_0\sqrt{\epsilon_0}}\;\curl \vb{u}_{\vb{m}}(\vb{x}),\lbl{HEexpansion2}
  \end{align}
  where 
\begin{align}
 q_{\vb{m}}''(t)=-\omega_{\vb{m}}q_{\vb{m}}(t),\lbl{HarmonicOscillator2}
\end{align}
  and where now, of course, the functions $\{\vb{u}_{\vb{m}}\}$ are solutions to the more complicated eigenvalue problem \rf{EigenValueProblem2}. In Appendix A we show that also for this more complicated situation the total energy of the system can be written in the form
\begin{align}
\mathcal{H}=\sum_{\vb{m}}\frac{1}{2}\left(\dot q_{\vb{m}}^2+\omega_{\vb{m}}^2q_{\vb{m}}^2\right).\lbl{OscillatorExpansion1}
\end{align}
In orther to arrive at this expression for the total energy we must use both the boundary conditions at the outer surface, $\Gamma_0$, as for the empty cavity, but also use the boundary conditions at the internal boundaries $\Gamma_j,\;\;j=1,...p$.
  
  The Heisenberg, Born, Jordan quantization now proceeds by first  introducing the canonical momentum variables
  \begin{align}
  p_{\vb{m}}=\dot q_{\vb{m}},\nonumber
  \end{align}
 so that  the Hamiltonian take the form
  \begin{align}
\mathcal{H}=\sum_{\vb{m}}\mathcal{H}_{\vb{m}},\lbl{OscillatorExpansion2}
  \end{align}
where 
\begin{align}
\mathcal{H}_{\vb{m}}=\frac{1}{2}\left(p_{\vb{m}}^2+\omega_{\vb{m}}^{2}q_{\vb{m}}^2\right).\lbl{SingleOscillator}
 \end{align}
The expression \rf{OscillatorExpansion2} is now the Hamiltonian for an uncoupled system of independent harmonic oscillators, one oscillator \rf{SingleOscillator} for each mode index $\vb{m}$.  We now perform canonical quantization to each of these harmonic oscillators by letting 
 \begin{align}
 q_{\vb{m}}\longrightarrow\hat q_{\vb{m}},\nonumber\\
 p_{\vb{m}}\longrightarrow\hat p_{\vb{m}},\nonumber
 \end{align}
 and imposing  the bosonic commutation rules on the operators $\hat q_{\vb{m}}$ and $\hat p_{\vb{m}}$
 \begin{align}
 \left[\hat q_{\vb{m}},\hat p_{\vb{m}'}\right]&=i\hbar\delta_{\vb{m},\vb{m}'},\nonumber\\
 \left[\hat q_{\vb{m}},\hat q_{\vb{m}'}\right]&=0,\nonumber\\
 \left[\hat p_{\vb{m}},\hat p_{\vb{m}'}\right]&=0.\lbl{HarmonicOscillatorCommutationRules}
  \end{align}
The Hamiltonian operators corresponding to the classical Hamiltonian \rf{OscillatorExpansion1} is now 
\begin{align}
\mathcal{\hat H}=\sum_{\vb{m}}\frac{1}{2}\left(\hat p_{\vb{m}}^2+\omega_{\vb{m}}^{2}\hat q_{\vb{m}}^2\right).\lbl{HOperator1}
\end{align}
\noindent We now introduce annihilation and creation operators for each harmonic oscillator 
\begin{align}
\hat a_{\vb{m}}&=\frac{1}{\sqrt{2\hbar\omega_{\vb{m}}}}\left(\omega_{\vb{m}}\hat q_{\vb{m}}+i\hat p_{\vb{m}}\right),\nonumber\\
\hat a_{\vb{m}}^{\dagger}&=\frac{1}{\sqrt{2\hbar\omega_{\vb{m}}}}\left(\omega_{\vb{m}}\hat q_{\vb{m}}-i\hat p_{\vb{m}}\right).\lbl{CreationAnnihilationOperators}
\end{align}
From the canonical commutation rules \rf{HarmonicOscillatorCommutationRules} it is easy to show that
\begin{align}
[\hat a_{\vb{m}},\hat a_{\vb{p}}^{\dagger}]&=\delta_{\vb{m}\vb{p}},\nonumber\\
[\hat a_{\vb{m}},\hat a_{\vb{p}}]&=0,\nonumber\\
[\hat a_{\vb{m}}^{\dagger},\hat a_{\vb{p}}^{\dagger}]&=0.\lbl{BosonicCommutationRules}
\end{align}
Inverting the identities \rf{CreationAnnihilationOperators} we have
\begin{align}
\hat q_{\vb{m}}&=\sqrt{\frac{\hbar}{2\omega_{\vb{m}}}}\left(\hat a_{\vb{m}}+\hat a_{\vb{m}}^{\dagger}\right),\nonumber\\
\hat p_{\vb{m}}&=i\sqrt{\frac{\hbar\omega_{\vb{m}}}{2}}\left(\hat a_{\vb{m}}-\hat a_{\vb{m}}^{\dagger}\right).\lbl{qInTermsOfAC}
\end{align}
Using the annihilation and creation operators  \rf{CreationAnnihilationOperators}, the expression for the total energy operator \rf{HOperator1} takes the form
\begin{align}
\mathcal{\hat H}=\sum_{\vb{m}}\;\hbar\;\omega_{\vb{m}}\;\mathcal{\hat N}_{\vb{m}},\lbl{EnergyOperator}
\end{align}
where 
\begin{align}
\mathcal{\hat N}_{\vb{m}}=\hat a_{\vb{m}}^{\dagger}\hat a_{\vb{m}},\lbl{NumberOperator}
\end{align}
is the number operator for the oscillator corresponding to index $\vb{m}$. In this formula we have dropped the infinite contribution from the vacuum energy.

It is easy to verify, using the Heisenberg equations of motion, that the time dependence of the  annihilation and creation operators are simply
\begin{align}
\hat a_{\vb{m}}(t)&=\hat a_{\vb{m}}\;e^{-i\;\omega_{\vb{m}}\; t},\nonumber\\
\hat a_{\vb{m}}^{\dagger}&=\hat a_{\vb{m}}^{\dagger}\;e^{i\;\omega_{\vb{m}}\; t}.\lbl{ACTimeDependence}
\end{align}
Using the identities \rf{ACTimeDependence} and \rf{qInTermsOfAC} in \rf{HEexpansion2} we find the following expressions for the quantized electric and magnetic field in the cavity
\begin{align}
\vb{\hat E}(\vb{x},t)&=-\sum_{\vb{m}}\;i\sqrt{\frac{\hbar\omega_{\vb{m}}}{2\epsilon_0}}\left(\hat a_{\vb{m}}e^{-i\;\omega_{\vb{m}}\; t}-\hat a_{\vb{m}}^{\dagger}e^{i\;\omega_{\vb{m}}\; t}\right)\;\vb{u}_{\vb{m}}(\vb{x}),\nonumber\\
\vb{\hat H}(\vb{x},t)&=\sum_{\vb{m}}\frac{1}{\mu_0}\sqrt{\frac{\hbar}{2\omega_{\vb{m}}\epsilon_0}}\left(\hat a_{\vb{m}}e^{-i\;\omega_{\vb{m}}\; t}+\hat a_{\vb{m}}^{\dagger}e^{i\;\omega_{\vb{m}}\; t}\right)\;\curl\vb{u}_{\vb{m}}(\vb{x}),\lbl{QuantumCavityFields}
\end{align}

\section{The non-interacting many-particle bosonic density operator}
For a non-interacting bosonic  many-body  system in thermal equilibrium at a temperature $T$, the maximum entropy principle tell us that the following density operator
\begin{align}
\hat\rho = \frac{1}{Z} \; \exp(-\frac{\mathcal{\hat H}}{kT}), \lbl{QMaxEntDensityOperator} 
\end{align}
is the one that in the best way represent the limited information
\begin{align}
 \left<  \mathcal{\hat H}\right>=E,\nonumber
 \end{align}
 that we have about the state of the many-body system. Using the expression for $\mathcal{\hat H}$ from the previous section we have
\begin{align}
\hat\rho = \frac{1}{Z} \; \exp(-\frac{1}{kT}\sum_{\vb{m}}\;\hbar\;\omega_{\vb{m}}\;\mathcal{\hat N}_{\vb{m}}). \lbl{BosonicDensityOperator} 
\end{align}
For any observable $\hat A$ for the bosonic system we compute the expectation value in the usual way
\begin{align}
\left<\hat A\right>&=Tr(\hat\rho\;\hat A)\nonumber\\
&=\sum_i\;<\psi_i\;| \hat\rho\;\hat A\;|\psi_i>,\nonumber
\end{align}
where the vectors $\psi_i$ is a basis for the state space. For the current situation the state space is the bosonic Fock-space. The basis for this space that we will use is the occupation number basis. The standard notation for this basis appear to us to be unwieldy and overpowering. We therefore introduce a different notation that is simpler and more compact.

In the current case and other cases that we will be interested in, the complete set of electromagnetic modes are indexed by a countable discrete set of indices, $I$, whose elements will be denoted by $\vb{m},\vb{n},\vb{p}$ etc. We use fat type for the indices since most of the time the set of  indices forms a countable subset of $\mathbb{Z}^r$ for some natural number $r$. Let us denote the set of finite subsets of the index set, I, by $\mathcal{P}(I)$
\begin{align}
\mathcal{P}(I)=\{U\subset I\;|\;| U|<\infty\}.
\end{align}
For any set in $S\in\mathcal{P}(I)$, define 
\begin{align}
\mathcal{F}(S)=\{n:I\rightarrow\mathbb{N}\cup\{0\}\;|\;n(\vb{p})\neq\;0\;\Leftrightarrow\;\vb{p}\in\;S\}.\nonumber
\end{align}
Thus, $\mathcal{F}(S)$ is the set of non-negative, integer valued functions with support $S$ defined on the index set $I$. We are now ready to describe the occupation number basis for the many-body Bosonic Fock space. Denoting the orthonormal basis by $\mathcal{B}$ we have
\begin{align}
\mathcal{B}=\{\left|n\right>\;|\;n\in\mathcal{F}(S),\;S\in\mathcal{P}(I)\;\}\lbl{FockBasis}
\end{align}
Thus any state in the many-body bosonic Fock space can be written as
\begin{align}
\left|\psi\right>=\sum_{S\in\mathcal{P}(I)}\sum_{n\in\mathcal{F}(S)}\;\;a_{S,n}\;\left|n\right>.\lbl{FockBasisExpansion}
\end{align}
This represents an expansion of the state vector $\left|\psi\right>$ into a countable sum of basis vectors where each basis vector is determined by a finite subset of indices such that the corresponding set of oscillators are above the ground state. Observe that a state vector $\left|n\right>$ depends on the $\text{\it set}$ of mode indices $S$ that defines the domain of the function $n$. Since a set, by definition, has no order assumed among it's elements it follows that the state vector \rf{FockBasisExpansion} is fully symmetric with respect to permutation of the oscillator indices $\vb{m}$. In the usual notation for state vectors one first assume an ordering of the indices $\vb{m}_1,\vb{m}_2,...$, write down an expression for the state vectors of the form $\left|n_{\vb{m}_1},n_{\vb{m}_2},..\right>$ and then remove the order by imposing permutation symmetry $\left|n_{\vb{m}_{\sigma(1)}},n_{\vb{m}_{\sigma(2)}},..\right>=\left|n_{\vb{m}_1},n_{\vb{m}_2},..\right>$. Here $\sigma$ is an arbitrary permutation of the natural numbers.

The occupation number basis is orthonormal
\begin{align}
\left<n\;|\;n'\right>=\delta_{n,n'}.\lbl{Normalization}
\end{align}
Completeness is expressed in terms of the identity
\begin{align}
\sum_{S\in\mathcal{P}(I)}\sum_{n\in\mathcal{F}(S)}\;\left|n\right>\left<n\right|=\vb{1}.
\end{align}
We will in the following find it useful to introduce a special type of Kronecker delta
\begin{align}
\delta_{\vb{m},S}=\left\{ 
\begin{tabular}{l}
1 \ \ $\mathbf{m}\in S$ \\ 
0 \ \ $\mathbf{m}\notin S$
\end{tabular}.
\right. 
\end{align}
For each $\vb{m}\in I$ it is also useful to define a function $\delta_{\vb{m}}:I\rightarrow\{0,1\}$ by
\begin{align}
\delta_{\vb{m}}(\mathbf{l})=\left\{ 
\begin{tabular}{l}
1 \ \ $\mathbf{m}=\mathbf{l}$ \\ 
0 \ \ $\mathbf{m}\neq\mathbf{l}$
\end{tabular}
\right. \;.
\end{align}
The action of the annihilation and creation operators are then given by 
\begin{align}
\hat a_{\vb{m}}\;\left|n\right>&=\delta_{\vb{m},S}\;\sqrt{ n(\vb{m})}\;\left|n-\delta_{\vb{m}}\right>,\nonumber\\
\hat a_{\vb{m}}^{\dagger}\;\left|n\right>&=\sqrt{ n(\vb{m})+1}\;\left|n+\delta_{\vb{m}}\right>.\lbl{CreationAnnihilationAction}
\end{align}
For the number operators $\mathcal{\hat N}_{\vb{m}}$ we then as expected have
\begin{align}
\mathcal{\hat N}_{\vb{m}}\;\left|n\right>&=\hat a_{\vb{m}}^{\dagger}\hat a_{\vb{m}}\;\left|n\right>=\delta_{\vb{m},S}\;\sqrt{ n(\vb{m})}\hat a_{\vb{m}}^{\dagger}\;\left|n-\delta_{\vb{m}}\right>\nonumber\\
&=\delta_{\vb{m},S}\; n(\vb{m})\;\left|n\right>.\lbl{NumberOperatorAction}
\end{align}

\section{Thermodynamical formalism for electromagnetic fields in a general cavity.}
In the previous sections of this document we have developed the mathematical tools that we need to investigate the thermodynamics of electromagnetic fields for a large class of cavities.In this section we use these tools to calculate thermodynamical quantities and investigate relations between them.

\subsection{Direct verification of the thermodynamical formalism.}
 Let us start with finding an expression for the partition function under the assumption that the only observable is the total energy. Using the notation from the previous section and the expression for the density operator for a non-interacting many-body bosonic system \rf{BosonicDensityOperator} we have
\begin{align}
Z&=Tr(\hat\rho)\nonumber\\
&=\sum_{S\in\mathcal{P}(I)}\sum_{n\in\mathcal{F}(S)}\;\left<n\right|\; \exp{-\frac{1}{kT}\sum_{\vb{l}\in I}\;\hbar\;\omega_{\vb{l}}\;\mathcal{\hat N}_{\vb{l}}}\;\left|n\right>\nonumber\\
&=\sum_{S\in\mathcal{P}(I)}\sum_{n\in\mathcal{F}(S)}\;\left<n\right|\;\Pi_{\vb{l}\in I}\exp{-\frac{\hbar \omega_{\vb{l}}}{kT}\mathcal{\hat N}_{\vb{l}}}\left|n\right>\nonumber\\
&=\sum_{S\in\mathcal{P}(I)}\sum_{n\in\mathcal{F}(S)}\;\left<n\right|\;\Pi_{\vb{l}\in I}\exp{-\frac{\hbar \omega_{\vb{l}}\delta_{\vb{l},S}n(\vb{l})}{kT}}\left|n\right>\nonumber\\
&=\sum_{S\in\mathcal{P}(I)}\sum_{n\in\mathcal{F}(S)}\;\Pi_{\vb{l}\in I}\exp{-\frac{\hbar \omega_{\vb{l}}\delta_{\vb{l},S}n(\vb{l})}{kT}}\nonumber\\
&=\sum_{S\in\mathcal{P}(I)}\sum_{n\in\mathcal{F}(S)}\;\Pi_{\vb{l}\in S}\exp{-\frac{\hbar \omega_{\vb{l}}n(\vb{l})}{kT}}.\lbl{PartitionFunctionFormula1}
\end{align}
Recall that the system we are discussing is contained in a cavity defined by a closed surface $\Gamma$ in $\mathbb{R}^3$. Thus, the operator for the total energy depends on this surface and so does the partition function. From the expression \rf{PartitionFunctionFormula1}, that we just derived, it is clear that the partition function depend in an implicit way on the surface through the spectrum, $\omega_{\vb{m}}$,  of eigenvalues of the operator $L=\curl(\curl\;)$ with the appropriate boundary conditions. Thus, in terms of direct variable dependence, the partition function is of the form
\begin{align}
Z=Z(T,\{\omega_{\vb{m}}\}_{m\in I})
\end{align}
Let us next calculate an expression for the expected number of photons in the electromagnetic mode indexed by $\vb{m}\in I$.
\begin{align}
N_{\vb{m}}&\equiv\left<\mathcal{\hat N}_{\vb{m}}\right>=Tr(\rho\;\mathcal{\hat N}_{\vb{m}})\nonumber\\
&=\sum_{S\in\mathcal{P}(I)}\sum_{n\in\mathcal{F}(S)}\;\left<n\right|\;\frac{1}{Z}\; \exp{-\frac{1}{kT}\sum_{\vb{l}\in I}\;\hbar\;\omega_{\vb{l}}\;\mathcal{\hat N}_{\vb{l}}}\;\mathcal{\hat N}_{\vb{m}}\;\left|n\right>\nonumber\\
&=\frac{1}{Z}\sum_{S\in\mathcal{P}(I)}\sum_{n\in\mathcal{F}(S)}\;\left<n\right|\;\Pi_{\vb{l}\in I}\exp{-\frac{\hbar \omega_{\vb{l}}}{kT}\mathcal{\hat N}_{\vb{l}}}\mathcal{\hat N}_{\vb{m}}\;\left|n\right>\nonumber\\
&=\frac{1}{Z}\sum_{S\in\mathcal{P}(I)}\sum_{n\in\mathcal{F}(S)}\;\delta_{\vb{m},S}\;n(\vb{m})\;\left<n\right|\;\Pi_{\vb{l}\in I}\exp{-\frac{\hbar \omega_{\vb{l}}}{kT}\mathcal{\hat N}_{\vb{l}}}\;\left|n\right>\nonumber\\
&=\frac{1}{Z}\sum_{S\in\mathcal{P}(I)}\sum_{n\in\mathcal{F}(S)}\;\delta_{\vb{m},S}\;n(\vb{m})\;\Pi_{\vb{l}\in I}\exp{-\frac{\hbar \omega_{\vb{l}}\delta_{\vb{l},S}n(\vb{l})}{kT}}\nonumber\\
&=\frac{1}{Z}\sum_{S\in\mathcal{P}(I)}\sum_{n\in\mathcal{F}(S)}\;\delta_{\vb{m},S}\;n(\vb{m})\;\Pi_{\vb{l}\in S}\exp{-\frac{\hbar \omega_{\vb{l}}n(\vb{l})}{kT}}.\lbl{MeanOfNumberOperator1}
\end{align}
This is a rather complicated expression, but observe that
\begin{align}
-\frac{k\;T}{\hbar}\frac{1}{Z}\frac{\partial\;Z}{\partial\;\omega_{\vb{m}}}&=\nonumber\\
&=-\frac{k\;T}{\hbar}\frac{1}{Z}\sum_{S\in\mathcal{P}(I)}\sum_{n\in\mathcal{F}(S)}\;\frac{\partial}{\partial_{\omega_{\vb{m}}}}\left(\Pi_{\vb{l}\in S}\exp{-\frac{\hbar \omega_{\vb{l}}n(\vb{l})}{kT}}\right)\nonumber\\
&=-\frac{k\;T}{\hbar}\frac{1}{Z}\sum_{S\in\mathcal{P}(I)}\sum_{n\in\mathcal{F}(S)}\nonumber\\
&\left(-\frac{\hbar}{kT}\right)\delta_{\vb{m},S}\;n(\vb{m})\;\Pi_{\vb{l}\in S}\exp{-\frac{\hbar \omega_{\vb{l}}n(\vb{l})}{kT}}\nonumber\\
&=\frac{1}{Z}\sum_{S\in\mathcal{P}(I)}\sum_{n\in\mathcal{F}(S)}\;\delta_{\vb{m},S}\;n(\vb{m})\;\Pi_{\vb{l}\in S}\exp{-\frac{\hbar \omega_{\vb{l}}n(\vb{l})}{kT}}.\lbl{MeanOfNumberOperator2}
\end{align}
Thus, we can conclude from \rf{MeanOfNumberOperator1} and \rf{MeanOfNumberOperator2} that
\begin{align}
N_{\vb{m}}\equiv\left<\mathcal{\hat N}_{\vb{m}}\right>=-\frac{kT}{\hbar}\frac{\partial\;\ln Z}{\partial\;\omega_{\vb{m}}}.\lbl{MeanOfNumberOperator}
\end{align}
This thermodynamic formula is a special case of the general variational formula \rf{QuantumFundamentalVariationalIdentity} for the case at hand where both the observable $\mathcal{\hat H}$ and the partition function $Z$ depends on a countable set of parameters $\{\omega_{\vb{m}}\}_{m\in I}$
\begin{align}
\frac{\partial\;\ln Z}{\partial\;\omega_{\vb{m}}}&=-\frac{1}{kT}\left<\frac{\partial \mathcal{\hat H}}{\partial\;\omega_{\vb{m}}}\right>,\nonumber\\
&\Downarrow\nonumber\\
\frac{\partial\;\ln Z}{\partial\;\omega_{\vb{m}}}&=-\frac{1}{kT}\left<\frac{\partial }{\partial\;\omega_{\vb{m}}}\left(\sum_{\vb{l}}\;\hbar\;\omega_{\vb{l}}\;\mathcal{\hat N}_{\vb{l}}\right)\right>,\nonumber\\
&\Downarrow\nonumber\\
\frac{\partial\;\ln Z }{\partial\;\omega_{\vb{m}}}&=-\frac{\hbar}{kT}\left<\;\mathcal{\hat N}_{\vb{m}}\;\right>,\nonumber\\
&\Updownarrow\nonumber\\
\left<\mathcal{\hat N}_{\vb{m}}\right>&=-\frac{kT}{\hbar}\frac{\partial\;\ln Z}{\partial\;\omega_{\vb{m}}}.\nonumber
\end{align}
Using formula \rf{MeanOfNumberOperator} we get the following formula for the expected value of the total energy of the system
\begin{align}
E&=\left<\mathcal{\hat H}\right>=\sum_{\vb{m}}\hbar\;\omega_{\vb{m}}\left<\mathcal{\hat N}_{\vb{m}}\right>\nonumber\\
&=\sum_{\vb{m}}\hbar\;\omega_{\vb{m}}\left(-\frac{kT}{\hbar}\right)\frac{\partial\;\ln Z}{\partial\;\omega_{\vb{m}}}\nonumber\\
&=-kT\;\sum_{\vb{m}}\;\omega_{\vb{m}}\frac{\partial\;\ln Z}{\partial\;\omega_{\vb{m}}}.\lbl{MeanOfEnergy}
\end{align}
Observe that
\begin{align}
\frac{1}{Z}\frac{\partial\; Z}{\partial T}&=\frac{1}{Z}\sum_{S\in\mathcal{P}(I)}\sum_{n\in\mathcal{F}(S)}\;\partial_T\;\Pi_{\vb{l}\in S}\exp{-\frac{\hbar \omega_{\vb{l}}n(\vb{l})}{kT}}\nonumber\\
&=\frac{1}{Z}\sum_{S\in\mathcal{P}(I)}\sum_{n\in\mathcal{F}(S)}\;\sum_{\vb{m}\in I}\delta_{\vb{m},S}\;n(\vb{m})\left(\frac{\hbar\;\omega_{\vb{m}}}{kT^2}\right)\;\Pi_{\vb{l}\in S}\exp{-\frac{\hbar \omega_{\vb{l}}n(\vb{l})}{kT}}\nonumber\\
&=\frac{1}{Z}\sum_{\vb{m}\in I}\left(\frac{\hbar\;\omega_{\vb{m}}}{kT^2}\right)\sum_{S\in\mathcal{P}(I)}\sum_{n\in\mathcal{F}(S)}\;\delta_{\vb{m},S}\;n(\vb{m})\;\Pi_{\vb{l}\in S}\exp{-\frac{\hbar \omega_{\vb{l}}n(\vb{l})}{kT}},\lbl{EnergyMean1}
\end{align}
and from the \rf{MeanOfEnergy} we get
\begin{align}
E&=-kT\;\sum_{\vb{m}}\hbar\;\omega_{\vb{m}}\frac{\partial\;\ln Z}{\partial\;\omega_{\vb{m}}}\nonumber\\
&=-kT\;\frac{1}{Z}\sum_{\vb{m}}\hbar\;\omega_{\vb{m}}\sum_{S\in\mathcal{P}(I)}\sum_{n\in\mathcal{F}(S)}\;\frac{\partial}{\partial_{\omega_{\vb{m}}}}\left(\Pi_{\vb{l}\in S}\exp{-\frac{\hbar \omega_{\vb{l}}n(\vb{l})}{kT}}\right)\nonumber\\
&=-kT\;\frac{1}{Z}\sum_{\vb{m}}\hbar\;\omega_{\vb{m}}\sum_{S\in\mathcal{P}(I)}\sum_{n\in\mathcal{F}(S)}
\left(-\frac{\hbar}{kT}\right)\delta_{\vb{m},S}\;n(\vb{m})\;\Pi_{\vb{l}\in S}\exp{-\frac{\hbar \omega_{\vb{l}}n(\vb{l})}{kT}}\nonumber\\
&=kT^2\;\frac{1}{Z}\sum_{\vb{m}\in I}\left(\frac{\hbar\;\omega_{\vb{m}}}{kT^2}\right)\sum_{S\in\mathcal{P}(I)}\sum_{n\in\mathcal{F}(S)}\;\delta_{\vb{m},S}\;n(\vb{m})\;\Pi_{\vb{l}\in S}\exp{-\frac{\hbar \omega_{\vb{l}}n(\vb{l})}{kT}}.\lbl{EnergyMean2}
\end{align}
Comparing \rf{EnergyMean1} and \rf{EnergyMean2} we get
\begin{align}
E=kT^2\frac{\partial\; \ln Z}{\partial T},\nonumber
\end{align}
which is the formula we get from the thermodynamical formalism.

\subsection{The partition function.}
For the simple situation of non-interacting bosons we can find an exact expression for the partition function. Starting with formula \rf{PartitionFunctionFormula1} we have
\begin{align}
Z&=\sum_{S\in\mathcal{P}(I)}\sum_{n\in\mathcal{F}(S)}\;\Pi_{\vb{l}\in S}\exp{-\frac{\hbar \omega_{\vb{l}}n(\vb{l})}{kT}}\nonumber\\
&=\Pi_{\vb{l}\in I}\sum_{S\in\mathcal{P}(I)}\sum_{n\in\mathcal{F}(S)}\;\exp{-\frac{\hbar \omega_{\vb{l}}\delta_{\vb{l},S}n(\vb{l})}{kT}}\nonumber\\
&=\Pi_{\vb{l}\in I}\sum_{N=0}^{\infty}
\underbrace{\sum_{S\in\mathcal{P}(I)}\sum_{n\in\mathcal{F}(S)}}_{n(\vb{l})=N}\;\exp{-\frac{\hbar \omega_{\vb{l}}\delta_{\vb{l},S}\;N}{kT}}\nonumber\\
&=\Pi_{\vb{l}\in I}\sum_{N=0}^{\infty}\alpha_{N,\vb{l}}\;\exp{-\frac{\hbar \omega_{\vb{l}}\;N}{kT}},\lbl{PartitionFunctionFormula2}
\end{align}
where we have defined 
\begin{align}
\alpha_{N,\vb{l}}&=\underbrace{\sum_{S\in\mathcal{P}(I)}\sum_{n\in\mathcal{F}(S)}}_{n(\vb{l})=N}\;1.\nonumber
\end{align}
All the numbers $\alpha_{N,\vb{l}}$ are obviously infinite and they are equal to the number of elements in certain subsets of the countable index set $I$. They are therefore all countable and thus all equal to the transfinite Cardinal number $\aleph_0$, the smallest number in Cantors infinite hierarchy of transfinite Cardinal numbers.

Using the arithmetic rules that apply to transfinite numbers we have
\begin{align}
Z&=\aleph_0\;\Pi_{\vb{l}\in I}\sum_{N=0}^{\infty}\;\exp{-\frac{\hbar \omega_{\vb{l}}\;N}{kT}}.\nonumber
\end{align}
In all applications of the partition function for computing means of observables, the same transfinite number $\aleph_0$ will occur in the numerator and denominator and they will therefore, again using the rules of transfinite arithmetic, cancel exactly. We can therefore disregard the transfinite multiplier in the formula for the partition function and thus get
\begin{align}
Z&=\Pi_{\vb{l}\in I}\sum_{N=0}^{\infty}\;\exp{-\frac{\hbar \omega_{\vb{l}}\;N}{kT}}=\Pi_{\vb{l}\in I}Z_{\vb{l}}(T,\omega_{\vb{l}}),\lbl{PartitionFunctionFormula3}
\end{align}
where
\begin{align}
Z_{\vb{l}}(T,\omega_{\vb{l}})&=\sum_{N=0}^{\infty}\;\exp{-\frac{\hbar \omega_{\vb{l}}\;N}{kT}}=\frac{1}{1-\exp{-\frac{\hbar\;\omega_{\vb{l}}}{kT}}}.\lbl{PartitionFunctionFormula4}
\end{align}

\subsection{Formulas for thermodynamical quantities pertaining to light fields in cavities.}
From the general thermodynamical formalism introduced in section two of these notes it is clear  that the partition function $Z=Z(\lambda_1,\cdots,\lambda_p\;\;\alpha_1,\cdots,\alpha_q)$ depends on the Lagrange multipliers $\{\lambda_i\}$, one for each observable, and possibly on a set of parameters $\{\alpha_j\}$, one for each fixed constraint that define the thermodynamical system  under investigation. Differentiation of $\ln Z$ with respect to the Lagrange multipliers determine all statistical moments of the observables, and differentiation with respect to the parameters determine other  important physical quantities through the general thermodynamic identities \rf{DParameterZ}, for the case of a countable number of parameters, and \rf{FundamentalVariationalIdentity} for the case of an uncountable number of parameters,for example a confining surface. We have in section \rf{pressure} seen how a formula like this  for the pressure is derived  from \rf{FundamentalVariationalIdentity}. For the calculation of all other physical quantities pertaining to the system we must do it by taking traces over the bosonic density operator \rf{QMaxEntDensityOperator}. We will see an example of this in the next section. In either case we end up with formulas that expresses all physical quantities pertaining to a thermodynamical system in terms of mode sums.

\subsubsection{Formulas for arbitrary cavities expressed in terms of mode sums}
Observe that all quantities that are calculable directly from the partition function  actually only depends on $\ln Z$. From the previous section we have
\begin{align}
\ln Z&=\ln\left(\Pi_{\vb{m}\in I}Z_{\vb{m}}(T,\omega_{\vb{m}})\right)=\sum_{\vb{m}}\ln Z_{\vb{m}}(T,\omega_{\vb{m}})\nonumber\\
&=-\sum_{\vb{m}}\ln \left(1-\exp{-\frac{\hbar\;\omega_{\vb{m}}}{kT}}\right)\lbl{LogZ}
\end{align}
Using the expression \rf{LogZ} for $\ln Z$ we find the following mode sum for the mean of the  total electromagnetic field energy  in the cavity
\begin{align}
E&=kT^2\frac{\partial\; \ln Z}{\partial T}\nonumber\\
&=-kT^2\frac{\partial\; }{\partial T}\left(\sum_{\vb{m}}\ln \left(1-\exp{-\frac{\hbar\;\omega_{\vb{m}}}{kT}}\right)\right)\nonumber\\
&=\sum_{\vb{m}}\frac{\hbar\;\omega_{\vb{m}}}{\exp{\frac{\hbar\;\omega_{\vb{m}}}{kT}}-1}.
\end{align}
Note that this formula holds for {\it any} cavity whatsoever. For the mean photon number in the cavity we have
\begin{align}
N&=\left<\mathcal{\hat N}\right>=\sum_{\vb{m}}\left<\mathcal{\hat N_{\vb{m}}}\right>=-\frac{kT}{\hbar}\sum_{\vb{m}}\frac{\partial\;\ln Z}{\partial\;\omega_{\vb{m}}}\nonumber\\
&=\sum_{\vb{m}}\frac{1}{\exp{\frac{\hbar\;\omega_{\vb{m}}}{kT}}-1}.\lbl{MeanOfNumberOperator}
\end{align}
We can not use the general formula \rf{FundamentalVariationalIdentity3.5}
\begin{align}
\left<p\right>_{\Gamma}= kT\;\frac{\delta_\Gamma\ln Z}{d_{\Gamma}V},\lbl{VariationalFormulaForPressure}
\end{align}
to express the pressure as a mode sum. The reason for this is that we only have the explicit dependence of the partition function on the mode frequencies, not it's dependence on the closed surface $\Gamma$ defining the cavity. Deforming the surface will or will not change the mode frequencies depending on the precise nature of the deformation. There are in fact for some cavities deformations that leads to no change in the mode frequencies at all. These deformations are called isospectral and there is a whole field of mathematics dedicated to investigating such deformations. Thus, there is in general no simple relation between the mode frequencies and deformations and therefore \rf{VariationalFormulaForPressure} is useless for determining a formula for the pressure for general cavities.

This means that for general cavities the pressure is one of the quantities that can only be computed by taking  traces over the bosonic density operator. We now proceed to do this for the simplest case of an empty cavity. The approach we use for these simplest kind of cavities can also be used to treat the general type of cavities introduced in section three, but then delicate questions regarding which stress tensor is the correct one to use in the presence of material bodies must first be resolved. If a choice has been made the calculations will proceed in a similar way to the one below for empty cavities.

The correct quantum Maxwell stress tensor for an empty cavity is
\begin{align}
\hat T&=\epsilon_0\vb{\hat E}\vb{\hat E}+\mu_0\vb{\hat H}\vb{\hat H}-\frac{1}{2}I\left(\epsilon_0\vb{\hat E}^2+\mu_0\vb{\hat H}^2\right)\lbl{MaxwellTensor1}
\end{align}
Recall that the quantum electromagnetic field inside the cavity can be written in the form
\begin{align}
  \vb{\hat E}(\vb{x},t)&=-\sum_{\vb{m}}\frac{\hat p_{\vb{m}}(t)}{\sqrt{\epsilon_0}}\;\vb{u}_{\vb{m}}(\vb{x}),\lbl{QEExpansion}\\
  \vb{\hat H}(\vb{x},t)&=\sum_{\vb{m}}\frac{\hat q_{\vb{m}}(t)}{\mu_0\sqrt{\epsilon_0}}\;\curl \vb{u}_{\vb{m}}(\vb{x}),\lbl{QHExpansion}
  \end{align}
  where we have
\begin{align}
\hat q_{\vb{m}}&=\sqrt{\frac{\hbar}{2\omega_{\vb{m}}}}\left(\hat a_{\vb{m}}+\hat a_{\vb{m}}^{\dagger}\right),\lbl{QqDef}\\
\hat p_{\vb{m}}&=i\sqrt{\frac{\hbar\omega_{\vb{m}}}{2}}\left(\hat a_{\vb{m}}-\hat a_{\vb{m}}^{\dagger}\right).\lbl{QpDef}
\end{align}
From \rf{MaxwellTensor1} and \rf{QqDef},\rf{QpDef} we get
\begin{align}
\hat T&=\sum_{\vb{m}}\sum_{\vb{p}}\{\hat p_{\vb{m}}\hat p_{\vb{p}}\vb{u}_{\vb{m}}\vb{u}_{\vb{p}}+c^2\hat q_{\vb{m}}\hat q_{\vb{q}}\curl \vb{u}_{\vb{m}}\curl\vb{u}_{\vb{p}}\}\nonumber\\
&-\frac{1}{2}I\sum_{\vb{m}}\sum_{\vb{p}}\{\hat p_{\vb{m}}\hat p_{\vb{p}}\vb{u}_{\vb{m}}\cdot\vb{u}_{\vb{p}}+c^2\hat q_{\vb{m}}\hat q_{\vb{p}}\curl \vb{u}_{\vb{m}}\cdot\curl\vb{u}_{\vb{p}}\}\lbl{MaxwellStressTensor2}
\end{align}
In order to find the force acting on a boundary $\Gamma$ of the cavity, we need to consider the quantity
\begin{align}
-\vb{n}\cdot\hat T\;|_{\Gamma}.\nonumber
\end{align}
But
\begin{align}
\curl\vb{u}_{\vb{m}}\cdot\vb{n}\;|_{\Gamma}=0.\nonumber
\end{align}
Thus, using \rf{MaxwellStressTensor2} we have
\begin{align}
-\vb{n}\cdot\hat T\;|_{\Gamma}&=\frac{1}{2}\vb{n}\sum_{\vb{m}}\sum_{\vb{p}}\{\hat p_{\vb{m}}\hat p_{\vb{p}}\vb{u}_{\vb{m}}\cdot\vb{u}_{\vb{p}}\;|_{\Gamma}+c^2\hat q_{\vb{m}}\hat q_{\vb{p}}\curl \vb{u}_{\vb{m}}\cdot\curl\vb{u}_{\vb{p}}\;|_{\Gamma}\}\nonumber\\
&-\sum_{\vb{m}}\sum_{\vb{p}}\hat p_{\vb{m}}\hat p_{\vb{p}}\vb{u}_{\vb{m}}(\vb{u}_{\vb{p}}\cdot\vb{n})\;|_{\Gamma}.\lbl{ForceTensor1}
\end{align}
But we also have 
\begin{align}
\vb{u}_{\vb{m}}\;|_{\Gamma}=u_{\vb{m}}\;\vb{n},\nonumber
\end{align}
for some scalar function $u_{\vb{m}}$. We can therefore write the expression \rf{ForceTensor1} in the form
\begin{align}
-\vb{n}\cdot\hat T\;|_{\Gamma}&=\hat p(\vb{x})\;\vb{n},\nonumber
\end{align}
where $\hat p(\vb{x})$ is the operator representing the pressure at a point $\vb{x}$  on the boundary. It is given by the expression
\begin{align}
\hat p(\vb{x})&=\frac{1}{2}\sum_{\vb{m}}\sum_{\vb{p}}\{c^2\hat q_{\vb{m}}\hat q_{\vb{p}}\curl \vb{u}_{\vb{m}}(\vb{x})\cdot\curl\vb{u}_{\vb{p}}(\vb{x})\;|_{\Gamma}-\hat p_{\vb{m}}\hat p_{\vb{p}}u_{\vb{m}}(\vb{x})u_{\vb{p}}(\vb{x})\;|_{\Gamma}\}\nonumber
\end{align}
The operator, $\hat p$, representing the spatial average of the pressure over a part $S$ of the boundary ,$\Gamma$,  is then given by
\begin{align}
\hat p&=\frac{1}{2}\sum_{\vb{m}}\sum_{\vb{p}}\{c^2\hat q_{\vb{m}}\hat q_{\vb{p}}\alpha_{\vb{m}\vb{p}}-\hat p_{\vb{m}}\hat p_{\vb{p}}\beta_{\vb{m}\vb{p}}\},\nonumber
\end{align}
where the numbers $\alpha_{\vb{m}\vb{p}}$ and $\beta_{\vb{m}\vb{p}}$ are given by 
\begin{align}
\alpha_{\vb{m}\vb{p}}&=\frac{1}{A({S})}\int_{S}dA\;\curl\vb{u}_{\vb{m}}\cdot\curl\vb{u}_{\vb{p}}.\nonumber\\
\beta_{\vb{m}\vb{p}}&=\frac{1}{A(S)}\int_{S}dA\;u_{\vb{m}}u_{\vb{p}}=\frac{1}{A(S)}\int_{S}dA\;\vb{u}_{\vb{m}}\cdot\vb{u}_{\vb{p}},\nonumber\\
\end{align}
The spatial average of the pressure over a part $S$ of a closed boundary $\Gamma$  enclosing a thermal electromagnetic field is then
\begin{align}
p&=\frac{1}{2}\sum_{\vb{m}}\sum_{\vb{p}}\{c^2\left<\hat q_{\vb{m}}\hat q_{\vb{p}}\right>\alpha_{\vb{m}\vb{p}}-\left<\hat p_{\vb{m}}\hat p_{\vb{p}}\right>\beta_{\vb{m}\vb{p}}\}.\lbl{MeanPressure1}
\end{align}
This is the promised mode sum formula for the pressure. A mode sum formula for the electromagnetic energy density in an empty cavity can be found in an entirely similar way to the above calculation.  Recall that the energy density of the electromagnetic field is given by the expression
\begin{align}
\hat u&=\frac{1}{2}\left(\epsilon_0\vb{\hat E}^2+\mu_0\vb{\hat H}^2\right).\lbl{EnergyDensity}
\end{align}
Using the exact same approach as for the pressure we find that the spatial average of the energy density in the cavity is 
\begin{align}
u&=\frac{1}{2}\sum_{\vb{m}}\sum_{\vb{p}}\{c^2\left<\hat q_{\vb{m}}\hat q_{\vb{p}}\right>\alpha_{\vb{m}\vb{p}}'+\left<\hat p_{\vb{m}}\hat p_{\vb{p}}\right>\beta_{\vb{m}\vb{p}}'\},\lbl{MeanEnergyDensity1}
\end{align}
where 
\begin{align}
\alpha_{\vb{m}\vb{p}}'&=\frac{1}{V(D)}\int_{D}dA\;\curl\vb{u}_{\vb{m}}\cdot\curl\vb{u}_{\vb{p}}.\nonumber\\
\beta_{\vb{m}\vb{p}}'&=\frac{1}{V(D)}\int_{D}dA\;\vb{u}_{\vb{m}}\cdot\vb{u}_{\vb{p}},\nonumber\\
\end{align}
We now need to calculate the expectation values occurring in the formulas for the pressure and the energy density. Let us first consider the expectation value $\left<\hat q_{\vb{m}}\hat q_{\vb{p}}\right>$. Using the formula 
\begin{align}
\hat q_{\vb{m}}&=\sqrt{\frac{\hbar}{2\omega_{\vb{m}}}}\left(\hat a_{\vb{m}}+\hat a_{\vb{m}}^{\dagger}\right),\nonumber
\end{align}
we have 
\begin{align}
\left<\hat q_{\vb{m}}\hat q_{\vb{p}}\right>=&-\frac{1}{2}\hbar\left(\omega_{\vb{m}}\omega_{\vb{p}}\right)^{-\frac{1}{2}}
(\left<\hat a_{\vb{m}}\hat a_{\vb{p}}\right>e^{-i\left(\omega_{\vb{m}}+\omega_{\vb{p}}\right)t}+\left<\hat a_{\vb{m}}\hat a_{\vb{p}}^{\dagger}\right>e^{-i\left(\omega_{\vb{m}}-\omega_{\vb{p}}\right)t}\nonumber\\
&+\left<\hat a_{\vb{m}}^{\dagger}\hat a_{\vb{p}}\right>e^{i\left(\omega_{\vb{m}}-\omega_{\vb{p}}\right)t}+\left<\hat a_{\vb{m}}^{\dagger}\hat a_{\vb{p}}^{\dagger}\right>e^{i\left(\omega_{\vb{m}}+\omega_{\vb{p}}\right)t}).\lbl{qqMean}
\end{align}
Let us calculate the expectation value $\left<\hat a_{\vb{m}}\hat a_{\vb{p}}^{\dagger}\right>$ explicitly. We have
\begin{align}
\left<\hat a_{\vb{m}}\hat a_{\vb{p}}^{\dagger}\right>&=Tr(e^{-\frac{\mathcal H}{kT}}\hat a_{\vb{m}}\hat a_{\vb{p}}^{\dagger})\nonumber\\
&=\sum_{S\in\mathcal{P}(I)}\sum_{n\in\mathcal{F}(S)}\;<n|\; \exp{-\frac{1}{kT}\sum_{\vb{l}\in I}\;\hbar\;\omega_{\vb{l}}\;\mathcal{\hat N}_{\vb{l}}}\hat a_{\vb{m}}\hat a_{\vb{p}}^{\dagger}|n>\nonumber\\
&=\sum_{S\in\mathcal{P}(I)}\sum_{n\in\mathcal{F}(S)}\; \exp{-\frac{1}{kT}\sum_{\vb{l}\in I}\;\hbar\;\omega_{\vb{l}}\; n(\vb{l})}<n|\;\hat a_{\vb{m}}\hat a_{\vb{p}}^{\dagger}|n>,\nonumber
\end{align}
and 
\begin{align}
<n|\;\hat a_{\vb{m}}\hat a_{\vb{p}}^{\dagger}|n>&=(n(\vb{p})^{\frac{1}{2}}<n|\;\hat a_{\vb{m}}|n+\delta_{\vb{p}}>\nonumber\\
&=(n(\vb{p})^{\frac{1}{2}}\delta_{\vb{m},S\cup\{\vb{p}\}}((n+\delta_{\vb{p}})(\vb{m})^{\frac{1}{2}}(\vb{p})^{\frac{1}{2}}\nonumber\\
&<n|\;\hat a_{\vb{m}}|n+\delta_{\vb{p}}-\delta_{\vb{m}}>=(n(\vb{m})+1)\delta_{\vb{m}\vb{p}}.\nonumber
\end{align}
Thus
\begin{align}
\left<\hat a_{\vb{m}}\hat a_{\vb{p}}^{\dagger}\right>&=\delta_{\vb{m}\vb{p}}\sum_{S\in\mathcal{P}(I)}\sum_{n\in\mathcal{F}(S)}\; (n(\vb{m})+1)\exp{-\frac{1}{kT}\sum_{\vb{l}\in I}\;\hbar\;\omega_{\vb{l}}\; n(\vb{l})}\nonumber\\
&=(N_{\vb{m}}+1)\delta_{\vb{m}\vb{p}}\lbl{Meanaadagger}
\end{align}
where $N_{\vb{m}}$ is the mean number of photons at frequency $\omega_{\vb{m}}$. In an entirely similar way we find 
\begin{align}
\left<\hat a_{\vb{m}}^{\dagger}\hat a_{\vb{p}}\right>&=N_{\vb{m}}\delta_{\vb{m}\vb{p}},\nonumber\\
\left<\hat a_{\vb{m}}\hat a_{\vb{p}}\right>&=0,\nonumber\\
\left<\hat a_{\vb{m}}^{\dagger}\hat a_{\vb{p}}^{\dagger}\right>&=0.\lbl{OtherMeans}
\end{align}
Inserting \rf{Meanaadagger} and \rf{OtherMeans} into \rf{qqMean} we finally get
\begin{align}
\left<\hat q_{\vb{m}}\hat q_{\vb{p}}\right>=\hbar\omega_{\vb{m}}^{-1}\left(N_{\vb{m}}+\frac{1}{2}\right)\delta_{\vb{m}\vb{p}}.\lbl{qqMean}
\end{align}
In an entirely similar way we get
\begin{align}
\left<\hat p_{\vb{m}}\hat p_{\vb{p}}\right>=\hbar\omega_{\vb{m}}\left(N_{\vb{m}}+\frac{1}{2}\right)\delta_{\vb{m}\vb{p}}.\lbl{ppMean}
\end{align}
Using these \rf{qqMean} and \rf{ppMean} in the expressions for the average pressure \rf{MeanPressure1}  and average energy density \rf{MeanEnergyDensity1}, we get the following mode sum formulas for these quantities
\begin{align}
p&=\frac{1}{2}\hbar\sum_{\vb{m}}(N_{\vb{m}}+\frac{1}{2})\{c^2\omega_{\vb{m}}^{-1}\alpha_{\vb{m}}-2\omega_{\vb{m}}\beta_{\vb{m}}\}\nonumber\\
u&=\frac{1}{2}\hbar\sum_{\vb{m}}(N_{\vb{m}}+\frac{1}{2})\{c^2\omega_{\vb{m}}^{-1}\alpha_{\vb{m}}'+\omega_{\vb{m}}\beta_{\vb{m}}'\}\nonumber\\,\lbl{MeanEnergyDensity2}
\end{align}
where $\alpha_{\vb{m}}$ is short for $\alpha_{\vb{m}\vb{m}}$ etc. and $\alpha(\omega_{\vb{m}})=\alpha_{\vb{m}}$ etc.
\subsubsection{Formulas for large cavities expressed in terms of mode integrals}
Let us start by being a little formal and introduce the general mode counting function  
\begin{align}
\mathcal{N}(\omega)=\sum_{\{\vb{m}\;|\;\omega_{\vb{m}}\leq\omega\}}\;1,\lbl{ModeCountingFunction}
\end{align}
and the associated mode density
\begin{align}
D(\omega)=\frac{d\;N(\omega)}{d\;\omega}.\lbl{ModeDensityMeasure}
\end{align}
Observe that in general the mode density is not an ordinary function but rather a generalized function. It is in fact a sum of delta functions. Thinking in terms of Lebesgue integrals all the mode sum formulas we have found can equivalently be written in terms of mode integrals 
\begin{align}
\ln Z&=-\int_0^{\infty}d\omega D(\omega)\;\ln \left(1-\exp{-\frac{\hbar\;\omega}{kT}}\right),\lbl{LnZ1}\\
E&=\int_0^{\infty}d\omega D(\omega)\;\frac{\hbar\;\omega}{\exp{\frac{\hbar\;\omega}{kT}}-1},\lbl{E11}\\
N&=\sum_{\vb{m}}\frac{1}{\exp{\frac{\hbar\;\omega_{\vb{m}}}{kT}}-1},\lbl{N1}\\
p&=\frac{1}{2}\int_0^{\infty}D(\omega)(\frac{1}{\exp{\frac{\hbar\;\omega}{kT}}-1}+\frac{1}{2})\{c^2\omega^{-1}\alpha(\omega)-2\omega\beta(\omega)\},\lbl{p1}\\
u&=\frac{1}{2}\int_0^{\infty}D(\omega)(\frac{1}{\exp{\frac{\hbar\;\omega}{kT}}-1}+\frac{1}{2})\{c^2\omega^{-1}\alpha'(\omega)+\omega\beta'(\omega)\}.\lbl{u1}
\end{align}
So far we have not done anything of significance, the mode sums and integrals are equivalent. The hard work remaining is to find approximations to the mode density, $D(\omega)$, in the large cavity limit. In these notes we will do this for two types cavities only. What needs to be done in more general situations should be clear from the two cases that we treat.

\subsection{Special cavities}
\subsubsection{Empty rectangular cavity with perfectly reflecting walls} \lbl{EmptyCavity}
Let us consider the specific case of a rectangular cavity whose sides have length $L_x,L_y$ and $L_z$. For these cavities the modes are indexed by $\vb{m}=(\vb{n},\sigma)$, where $\vb{n}=(n_x,n_y,n_z)$ are 3D positive nonzero  integer vectors and where $\sigma\in\{+,-\}$ signify the polarization of the mode. The derivation and general formulas for the modes can be found in \cite{Masud1}. The dispersion relation for modes of both types of polarizations is
\begin{align}
\left(\frac{\omega}{c}\right)^2=\frac{\pi^2n_x^2}{L_x^2}+\frac{\pi^2n_y^2}{L_y^2}+\frac{\pi^2n_z^2}{L_z^2},\lbl{EmptyRectangularPerfectlyReflectingDiespersionRelation1}
\end{align}
where $n_x$,$n_y$ and $n_z$ are positive integers. For this case it is easy and well known how to find the mode density in the large cavity limit. We however include it here because we must repeat it in a less familiar situation in the next section. 

We can evidently write the dispersion relation in the form
\begin{align}
\frac{\omega}{c}=k_{\vb{n}},\lbl{EmptyRectangularPerfectlyReflectingDiespersionRelation1}
\end{align}
where $k_{\vb{n}}$ are the discrete wavenumbers labeled by  positive integer vector $\vb{n}=(n_x,n_y,n_z)$. Let us introduce a three dimensional space of continuous wave vectors, $\vb{k}$ whose vector norm we write as $k=||\vb{k}||$.  The discrete wave vectors are then located on a square grid of equidistant points in the first quadrant of $\vb{k}$ space. Thus to each discrete wave vector and thus to each cavity mode we can assign a volume in wave number space given by
\begin{align}
\delta=\frac{\pi^3}{L_xL_yL_z}=\frac{\pi^3}{V},\lbl{Modevolume1}
\end{align}
where $V=L_xL_yL_z$ is the volume of the cavity.This formula shows that in the large cavity limit the volume taken up by each mode is arbitrarily small. The number of modes in the part of a spherical shell of radius $k$ and thickness $dk$ cut out by the first quadrant is then given by
\begin{align}
\pi k^2dk,
\end{align}
where we have taken into account the fact that there are two independent  polarizations of the electromagnetic field modes. The mode density then must be
\begin{align}
D(\omega)=\frac{\pi k^2dk,}{\delta}=\frac{V\omega^2}{\pi^2c^3},\lbl{ModeDensity1}
\end{align}
where we have used the dispersion relation to relate wave number to frequency. We insert the explicit formula for the mode density into \rf{E1} and find that the total energy is given by the explicit mode integral
\begin{align}
E&=V\;\int_0^{\infty}d\omega\;\mathcal{E}(\omega),\nonumber\\
\end{align}
where 
\begin{align}
\mathcal{E}(\omega)&=\frac{\hbar\omega^3}{\pi^2\;c^3\left(\exp{\frac{\hbar\omega}{kT}}-1\right)},\lbl{PlanckDistribution}
\end{align}
which we recognize as the Planck energy density spectrum in free space. The mode integrals for the natural logarithm of the partition function \rf{LnZ1}, the total energy\rf{E11} and the total particle number \rf{N1} can now be explicitly solved and we find the expressions
\begin{align}
\ln Z(T,V)&=\left(\frac{k^3\pi^2}{45c^3\hbar^3}\right)VT^3,\lbl{PartitionFunctionFormula6}\\
E(T,V)&=\left(\frac{k^4\pi^2}{15\hbar^3c^3}\right)VT^4,\lbl{TotalEnergy}\\
N(T,V)&=\left(\frac{2\;k^3\xi(3)}{\hbar^3c^3\pi^2}\right)VT^3,\lbl{TotalParticalNumber}
\end{align}
where $\xi(x)$ is the Riemann zeta function. 
The partition function is now an explicit function of the volume and the thermodynamic formula \rf{FundamentalVariationalIdentity4} for the pressure  yields
\begin{align}
p(T)&=kT\left(\frac{k^3\pi^2}{45c^3\hbar^3}\right)T^3.\lbl{ThermodynamicPressure}
\end{align}
Observe that the well known formula 
\begin{align}
p(T)=\frac{1}{3}u(T),\lbl{PressureEnergyLaw}
\end{align}
where $u(T)=E(T,V)/V$ is the energy density in the cavity, now appears. At this point it is worth recalling that we have different mode sum expressions for the pressure \rf{p1} and electromagnetic energy density \rf{u1} that holds for any empty cavity. In the large cavity limit these mode sums become mode integrals that should give the same expressions as the ones that we got from the general thermodynamic formalism. This is not at all evident and is thus an excellent test of the mode sum approach to $p$ and $u$.
In order to show that the corresponding expressions are the same, we first need to calculate the geometric quantities $\alpha_{\vb{m}},\beta_{\vb{m}},\alpha'_{\vb{m}}$ and $\beta'_{\vb{m}}$.
In order to do this calculation we will need the $\alpha_{\vb{m}}$ and $\beta_{\vb{m}}$ and for this we need explicit expressions for the electromagnetic modes in the cavity. For a rectangular cavity with perfectly reflecting walls it is straight forward to find these modes. Calculating the geometric quantities from the modes are more of an effort, but with a help from the symbolic algebra system Mathematica the integrals can be done and we get the following simple formulas
\begin{align}
c^2\omega_{\vb{m}}^{-1}\alpha_{\vb{m}}-\omega_{\vb{m}}\beta_{\vb{m}}&=\frac{2c^2k_{\vb{m}x}^2}{\pi V\omega_{\vb{m}}},\lbl{gamma2facesp1}\\
c^2\omega_{\vb{m}}^{-1}\alpha'_{\vb{m}}+\omega_{\vb{m}}\beta'_{\vb{m}}&=\frac{2\omega_{\vb{m}}}{V}.\lbl{delta2facesp1}
\end{align}
 Note that these expressions are the same for both polarizations.  From  \rf{gamma2facesp1}, \rf{delta2facesp1} we get the following expressions for the pressure, averaged over the whole surface, and the energy density.
\begin{align}
p&=\frac{1}{V}\sum_{\vb{m}}\frac{c^2\hbar k_{\vb{m}x}^2}{\pi V\omega_{\vb{m}}}(N_{\vb{m}}+\frac{1}{2}),\lbl{MeanPressure2faces1}\\
u&=\frac{1}{V}\sum_{\vb{m}}\hbar\omega_{\vb{m}}(N_{\vb{m}}+\frac{1}{2}).\lbl{MeanEnergyDensity4}
\end{align}
Because the mode sums depends not only on frequency, but also depends explicitly on a component of the wave vector, we will need a mode density that depends on both frequency and angles. Essentially we consider the number of modes in a volume in wave number space that is the part of the spherical shell in the first quadrant that has thickness $dk$ and angular spread $d\theta$ and $d\phi$ where $\theta$ is the angle with respect to the positive z-axis and $\phi$ the angle with respect to the positive x-axis  in a standard spherical coordinate system. Reasoning as earlier in this section we now find 
\begin{align}
D(\omega,\theta,\phi)=\frac{2V\omega^2\sin(\theta)}{\pi^3 c^3},
\end{align}
and the mode sums for the pressure and the electromagnetic energy density turns into the mode integrals
 \begin{align}
p&=\int_0^{\infty}\int_{0}^{\frac{\pi}{2}}  \int_{0}^{\frac{\pi}{2}}d\omega d\theta d\phi\frac{\hbar\omega^3}{c^3\pi^2}\frac{2\sin(\theta)^3\cos(\phi)^2}{\pi}(N_{\omega}+\frac{1}{2}),\nonumber\\
&=\int_0^{\infty}d\omega\frac{\hbar\omega^3}{3c^3\pi^2}(N_{\omega}+\frac{1}{2})\lbl{MeanPressure2faces2}\\
u&=\int_0^{\infty}\int_{0}^{\frac{\pi}{2}}  \int_{0}^{\frac{\pi}{2}}d\omega d\theta d\phi\frac{2\hbar\omega^3\sin(\theta)}{c^3\pi^3}(N_{\omega}+\frac{1}{2})\nonumber\\
&=\int_0^{\infty}d\omega\frac{\hbar\omega^3}{c^3\pi^2}(N_{\omega}+\frac{1}{2}),\lbl{MeanEnergyDensity5}
\end{align}
which even without solving the integrals give us the expected relation
\begin{align}
p=\frac{1}{3}u.\nonumber
\end{align}
If we drop the divegent vacuum contribution, $\frac{1}{2}$, from formulas \rf{MeanPressure2faces2} and \rf{MeanEnergyDensity5} for the pressure and energy it is straight forward to do the integrals and show that they do indeed coincide with the ones computed from the thermodynamical formalism.

\subsubsection{Periodic cavity containing a dielectric slab}
In this section we will study the thermodynamic properties of an electromagnetic field in a periodic cavity that contains a dielectric slab.  The goal is to find explicit expressions that turns mode sums into mode integral for this situation. 

  The cavity is oriented along the axes of a positively oriented coordinate system with the $z$ axis pointing vertically up. The sides of the cavity have lengths, $L_x$, $L_y$ and $L_z$. The slab is parallel to the $x-y$ plane and is centered around the point $z=a$ on the $z$ axis. The horizontal dimensions of the slab are $L_x$ and $L_y$ and thus fills the cavity horizontally. In the vertical direction the slab is contained in the interval $[a-b,a+b]$ and is thus of thickness $L=2b$. The cavity is periodic in all directions of period $L_x$,$L_y$ and $L_z$ and the slab is horizontally periodic with same period as the cavity, $L_x$ and $L_y$. We denote the domain containing the slab by $D$. The part of the cavity below the slab is denoted by $D^-$ and the part of the cavity above the slab is denoted by $D^+$. The refractive index $n(\vb{x})$ varies over the cavity and has the value $n$ inside the slab and $n_0$ under and over the slab. The unit normal along the $z$ axis is denoted by $\widehat{\vb{k}}$.

  For the cavity just described, the general mode equation for the electric field \rf{EigenValueProblem2} takes the form
\begin{align}
\laplacian\vb{E}^-&=-\left(\frac{\omega}{c}\right)n_0^2\vb{E}^- \nonumber\\
 \div\vb{E}^-&=0,\;\;\;\;\;\;\;\;\;\;\;\;\;\;\;\;\;\;\;\;\;\;\;\;\;\;\;\;\;\;\;\;\;\;\;\;\;\;\;\;\;\;\vb{x}\in D^-,\lbl{MinusEq}
\end{align}

\begin{align}
\laplacian\vb{E}&=-\left(\frac{\omega}{c}\right)n^2\vb{E}\nonumber\\
 \div\vb{E}&=0,\;\;\;\;\;\;\;\;\;\;\;\;\;\;\;\;\;\;\;\;\;\;\;\;\;\;\;\;\;\;\;\;\;\;\;\;\;\;\;\;\;\;\vb{x}\in D, \lbl{Eq}
\end{align}

\begin{align}
\laplacian\vb{E}^+&=-\left(\frac{\omega}{c}\right)n_0^2\vb{E}^+ \nonumber\\
 \div\vb{E}^+&=0,\;\;\;\;\;\;\;\;\;\;\;\;\;\;\;\;\;\;\;\;\;\;\;\;\;\;\;\;\;\;\;\;\;\;\;\;\;\;\;\;\;\;\vb{x}\in D^+.\lbl{PlussEq}
\end{align}

\noindent The magnetic field is under the assumptions given determined by the electric field through the relation
\begin{align}
\vb{H}=\frac{i}{\mu_0\omega}\curl{\vb{E}}.\lbl{HField}
\end{align}

\begin{figure}[t]
\centering
  \includegraphics[scale=0.4]{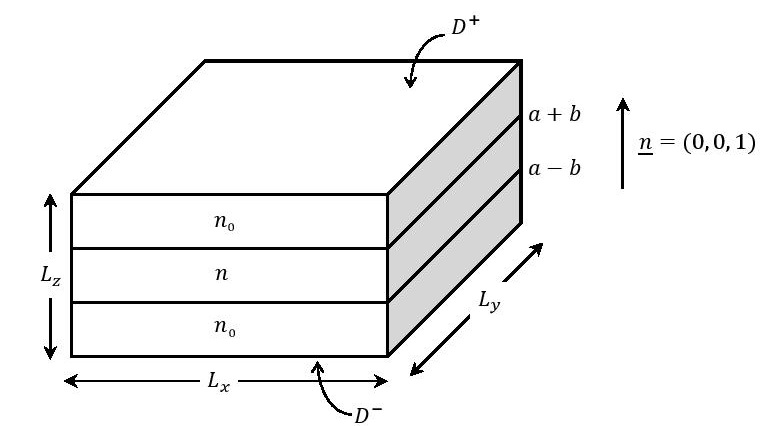}
  \caption{$ $}
  \label{fig1}
\end{figure}

\noindent Let us consider \rf{MinusEq} in detail. We look for solutions of the form
\begin{align}
\vb{E}(\vb{x},z)=\vb{E}^{-}e^{i\beta^{-}z}e^{i\vb{\xi}^{-}\cdot\vb{x}},\lbl{Em}
\end{align}
where $\vb{x}=(x,y)$ and $\vb{\xi}^{-}=(\xi_{x}^{-},\xi_{y}^{-})$.  In order for \rf{Em} to be a solution to \rf{MinusEq} we must have

\begin{align}
\left(\beta^{-}\right)^2&=\left(\frac{\omega}{c}\right)^2n_0^2-\left(\xi^{-}\right)^{2},\nonumber\\
&\beta^{-}e_z^{-}+\vb{\xi}^{-}\cdot\vb{e}^{-}=0,\lbl{EmParameterConstraints}
\end{align}
where we $e_z^{-}$ and $\vb{e}^{-}=\left(e^{-}_x,e^{-}_y\right)$ are defined through
\begin{align}
\vb{E}^{-}&=\left(\vb{e}^{-},e_z^{-}\right).
\end{align}
The solution in the domains $D$ and $D^{+}$ are treated in the same way. Thus in summary, the general solutions to the Maxwell equations  in all the three domains are 
\begin{align}
\vb{E}\left(\vb{x},z\right)&=\vb{E}_{1}^{-}e^{i\beta^{-}z}e^{i\vb{\xi}^{-}\cdot\vb{x}}\nonumber\\
&+\vb{E}_{2}^{-}e^{-i\beta^{-}z}e^{i\vb{\xi}^{-}\cdot\vb{x}},\;\;\;\;\;\;\;\;\vb{x}\in D^{-},\lbl{EMinus}\\
&\nonumber\\
\vb{E}\left(\vb{x},z\right)&=\vb{E}_{1}e^{i\beta z}e^{i\vb{\xi}\cdot\vb{x}}\nonumber\\
&+\vb{E}_{2}e^{-i\beta z}e^{i\vb{\xi}\cdot\vb{x}},\;\;\;\;\;\;\;\;\;\;\;\;\;\vb{x}\in D,\lbl{E}\\
&\nonumber\\
\vb{E}\left(\vb{x},z\right)&=\vb{E}_{1}^{+}e^{i\beta^{+}z}e^{i\vb{\xi}^{+}\cdot\vb{x}}\nonumber\\
&+\vb{E}_{2}^{+}e^{-i\beta^{+}z}e^{i\vb{\xi}^{+}\cdot\vb{x}},\;\;\;\;\;\;\;\;\vb{x}\in D^{+},\lbl{EPlus}
\end{align}
where
\begin{align}
\vb{E}^{-}&=\left(\vb{e}^{-},e_z^{-}\right),\nonumber\\
\vb{e}^{-}&=\left(e^{-}_x,e^{-}_y\right),\nonumber\\
\beta^{-}&=\sqrt{\left(\frac{\omega}{c}\right)^2n_0^2-\left(\xi^{-}\right)^{2}},\nonumber\\
\vb{\xi}^{-}&=(\xi_{x}^{-},\xi_{y}^{-}),\lbl{EMinusPar}\\
\vb{E}&=\left(\vb{e},e_z\right),\nonumber\\
\vb{e}&=\left(e_x,e_y\right),\nonumber\\
\beta&=\sqrt{\left(\frac{\omega}{c}\right)^2n^2-\left(\xi\right)^{2}},\nonumber\\
\vb{\xi}&=(\xi_{x},\xi_{y}),\lbl{EPar}\\
\vb{E}^{+}&=\left(\vb{e}^{+},e_z^{+}\right),\nonumber\\
\vb{e}^{+}&=\left(e^{+}_x,e^{+}_y\right),\nonumber\\
\beta^{+}&=\sqrt{\left(\frac{\omega}{c}\right)^2n_0^2-\left(\xi^{+}\right)^{2}},\nonumber\\
\vb{\xi}^{+}&=(\xi_{x}^{+},\xi_{y}^{+}).\lbl{EPlusPar}
\end{align}
Let us now start imposing boundary conditions on the solutions \rf{EMinus},\rf{E},\rf{EPlus}. 

\noindent Periodicity in the $z$ direction is ensured if 
\begin{align}
\vb{E}\left(\vb{x},0^{+}\right)&=\vb{E}\left(\vb{x},0^{-}\right)\lbl{zPeriodic1}\\
\partial_z\vb{E}\left(\vb{x},0^{+}\right)&=\partial_z\vb{E}\left(\vb{x},0^{-}\right).\lbl{zPeriodic2}
\end{align}
Inserting \rf{EMinus} and \rf{EPlus} into \rf{zPeriodic1} we get the equation
\begin{align}
&\vb{E}_{1}^{-}e^{i\vb{\xi}^{-}\cdot\vb{x}}+\vb{E}_{2}^{-}e^{i\vb{\xi}^{-}\cdot\vb{x}}\nonumber\\
&=\vb{E}_{1}^{+}e^{i\beta^{+}L_z}e^{i\vb{\xi}^{+}\cdot\vb{x}}+\vb{E}_{2}^{+}e^{-i\beta^{+}L_z}e^{i\vb{\xi}^{+}\cdot\vb{x}}.\lbl{zPeriodic1.1}
\end{align}
In order for this equation to have any solutions at all, we must clearly impose the condition
\begin{align}
\vb{\xi}^{-}=\vb{\xi}^{+},\lbl{xiPM}
\end{align}
and then \rf{zPeriodic1.1} reduces to 
\begin{align}
&\vb{E}_{1}^{-}+\vb{E}_{2}^{-}=\vb{E}_{1}^{+}e^{i\beta^{+}L_z}+\vb{E}_{2}^{+}e^{-i\beta^{+}L_z}.\lbl{zPeriodic1.2}
\end{align}
From the second boundary condition \rf{zPeriodic2} we get, using the identity \rf{xiPM}
\begin{align}
\beta^{-}\left(\vb{E}_{1}^{-}-\vb{E}_{2}^{-}\right)=\beta^{+}\left(\vb{E}_{1}^{+}e^{i\beta^{+}L_z}-\vb{E}_{2}^{+}e^{-i\beta^{+}L_z}\right).\lbl{zPeriodic2.1}
\end{align}
From \rf{EMinusPar} and \rf{EPlusPar}, using \rf{xiPM},  we conclude that $\beta^{+}=\beta^{-}$. But then it is easy to see that the only solution to \rf{zPeriodic1.2} and \rf{zPeriodic2.1} is 
\begin{align}
\vb{E}^{+}_1&=e^{-i\beta^{-}L_z}\vb{E}^{-}_1,\nonumber\\
\vb{E}^{+}_2&=e^{i\beta^{-}L_z}\vb{E}^{-}_2.\lbl{EplusEminus}
\end{align}
For the same reason that  \rf{zPeriodic1.1} enforces the condition \rf{xiPM}, we conclude that the electromagnetic interface conditions at the slab boundaries $z=a\pm b$ can only hold if 
\begin{align}
\vb{\xi}=\vb{\xi}^{+}=\vb{\xi}^{-}.\nonumber
\end{align}

Inserting what we have found so far into the expressions for the electric field in $D^{-},D$ and $D^{+}$, we have 
\begin{align}
\vb{E}\left(\vb{x},z\right)&=\vb{E}_{1}^{-}e^{i\beta^{-} z}e^{i\vb{\xi}\cdot\vb{x}}\nonumber\\
&+\vb{E}_{2}^{-}e^{-i\beta^{-} z}e^{i\vb{\xi}\cdot\vb{x}},\;\;\;\;\;\;\;\;\vb{x}\in D^{-},\lbl{EMinus1}\\
&\nonumber\\
\vb{E}\left(\vb{x},z\right)&=\vb{E}_{1}e^{i\beta z}e^{i\vb{\xi}\cdot\vb{x}}\nonumber\\
&+\vb{E}_{2}e^{-i\beta z}e^{i\vb{\xi}\cdot\vb{x}},\;\;\;\;\;\;\;\;\;\;\;\;\;\vb{x}\in D,\lbl{E1}\\
&\nonumber\\
\vb{E}\left(\vb{x},z\right)&=\vb{E}_{1}^{-}e^{i\beta^{-}\left( z-L_z\right)}e^{i\vb{\xi}\cdot\vb{x}}\nonumber\\
&+\vb{E}_{2}^{+}e^{-i\beta^{-} \left( z-L_z\right)}e^{i\vb{\xi}\cdot\vb{x}},\;\;\;\;\;\;\;\;\vb{x}\in D^{+}.\lbl{EPlus1}
\end{align}
Periodicity in the $x$ and $y$ directions is clearly ensured if 
\begin{align}
\vb{\xi}=\left(\frac{2\pi n_x}{L_x},\frac{2\pi n_y}{L_y}\right),\lbl{xiDef}
\end{align}
where $n_x$ and $n_y$ are arbitrary integers.
We now pose the boundary conditions at the boundaries of the slab located at $z=a\pm b$. 

\noindent Let us first consider the magnetic boundary condition. From \rf{EMinus1} and \rf{E1} we have
\begin{align}
\curl\vb{E}\left(\vb{x},z\right)&=i\left(\vb{\xi},\beta^{-}\right)\cross\vb{E}_{1}^{-}e^{i\beta^{-}z}e^{i\vb{\xi}\cdot\vb{x}}\nonumber\\
&+i\left(\vb{\xi},-\beta^{-}\right)\cross\vb{E}_{2}^{-}e^{-i\beta^{-}z}e^{i\vb{\xi}\cdot\vb{x}},\;\;\;\;\;\;\;\;\vb{x}\in D^{-},\lbl{CurlEMinus}\\
&\nonumber\\
\curl\vb{E}\left(\vb{x},z\right)&=i\left(\vb{\xi},\beta\right)\cross\vb{E}_{1}e^{i\beta z}e^{i\vb{\xi}\cdot\vb{x}}\nonumber\\
&+i\left(\vb{\xi},-\beta\right)\cross\vb{E}_{2}e^{-i\beta z}e^{i\vb{\xi}\cdot\vb{x}},\;\;\;\;\;\;\;\;\;\;\;\;\;\vb{x}\in D.\lbl{CurlE}\\
&\nonumber
\end{align}
Using the formula \rf{HField} expressing the magnetic field in terms of the curl of the electric field given above in \rf{CurlEMinus},\rf{CurlE}, the continuity of the magnetic field at $z=a-b$ is ensured if
\begin{align}
&\left(\vb{\xi},\beta^{-}\right)\cross\vb{E}_{1}^{-}e^{i\beta^{-}\left(a-b\right)}
+\left(\vb{\xi},-\beta^{-}\right)\cross\vb{E}_{2}^{-}e^{-i\beta^{-}\left(a-b\right)}\nonumber\\
&=\left(\vb{\xi},\beta\right)\cross\vb{E}_{1}e^{i\beta \left(a-b\right)}
+\left(\vb{\xi},-\beta\right)\cross\vb{E}_{2}e^{-i\beta \left(a-b\right)}.\lbl{MagBc1}
\end{align}
In a similar way, continuity of the magnetic field at the other boundary of the slab at $z=a+b$ is ensured if
\begin{align}
&\left(\vb{\xi},\beta\right)\cross\vb{E}_{1}e^{i\beta\left(a+b\right)}
+\left(\vb{\xi},-\beta\right)\cross\vb{E}_{2}e^{-i\beta\left(a+b\right)}\nonumber\\
&=\left(\vb{\xi},\beta^{-}\right)\cross\vb{E}_{1}^{-}e^{i\beta^{-} \left(a+b-L_z\right)}
+\left(\vb{\xi},-\beta^{-}\right)\cross\vb{E}_{2}^{-}e^{-i\beta^{-} \left(a+b-L_z\right)}.\lbl{MagBc2}
\end{align}
The continuity of the parallel component of $\vb{E}$ and the normal component of $\vb{D}$ at the slab boundaries give us the following equations
\begin{align}
&\vb{\hat k}\cross\vb{E}_{1}^{-}e^{i\beta^{-}\left(a-b\right)}
+\vb{\hat k}\cross\vb{E}_{2}^{-}e^{-i\beta^{-}\left(a-b\right)}\nonumber\\
&=\vb{\hat k}\cross\vb{E}_{1}e^{i\beta \left(a-b\right)}
+\vb{\hat k}\cross\vb{E}_{2}e^{-i\beta \left(a-b\right)},\lbl{ParallelEBc1}\\
&\nonumber\\
&n_0^2\;\vb{\hat k}\cdot\vb{E}_{1}^{-}e^{i\beta^{-}\left(a-b\right)}
+n_0^2\;\vb{\hat k}\cdot\vb{E}_{2}^{-}e^{-i\beta^{-}\left(a-b\right)}\nonumber\\
&=n^2\;\vb{\hat k}\cdot\vb{E}_{1}e^{i\beta \left(a-b\right)}
+n^2\;\vb{\hat k}\cdot\vb{E}_{2}e^{-i\beta \left(a-b\right)},\lbl{NormalDBc1}\\
&\nonumber\\
&\vb{\hat k}\cross\vb{E}_{1}e^{i\beta\left(a+b\right)}
+\vb{\hat k}\cross\vb{E}_{2}e^{-i\beta\left(a+b\right)}\nonumber\\
&=\vb{\hat k}\cross\vb{E}_{1}^{-}e^{i\beta^{-} \left(a+b-L_z\right)}
+\vb{\hat k}\cross\vb{E}_{2}^{-}e^{-i\beta^{-} \left(a+b-L_z\right)},\lbl{ParallelEBc2}\\
&\nonumber\\
&n^2\;\vb{\hat k}\cdot\vb{E}_{1}e^{i\beta\left(a+b\right)}
+n^2\;\vb{\hat k}\cdot\vb{E}_{2}e^{-i\beta\left(a+b\right)}\nonumber\\
&=n_0^2\;\vb{\hat k}\cdot\vb{E}_{1}^{-}e^{i\beta^{-} \left(a+b-L_z\right)}
+n_0^2\;\vb{\hat k}\cdot\vb{E}_{2}^{-}e^{-i\beta^{-} \left(a+b-L_z\right)}.\lbl{NormalDBc2}
\end{align}
Using \rf{EMinusPar},\rf{EPar} and \rf{EPlusPar}, these equations can be rewritten in the form
\begin{align}
&\vb{\hat k}\cross\vb{e}_{1}^{-}e^{i\beta^{-}\left(a-b\right)}
+\vb{\hat k}\cross\vb{e}_{2}^{-}e^{-i\beta^{-}\left(a-b\right)}\nonumber\\
&=\vb{\hat k}\cross\vb{e}_{1}e^{i\beta \left(a-b\right)}
+\vb{\hat k}\cross\vb{e}_{2}e^{-i\beta \left(a-b\right)},\lbl{ParallelEBc1.1}\\
&\nonumber\\
&n_0^2\;e_{z1}^{-}e^{i\beta^{-}\left(a-b\right)}
+n_0^2\;e_{z2}^{-}e^{-i\beta^{-}\left(a-b\right)}\nonumber\\
&=n^2\;e_{z1}e^{i\beta \left(a-b\right)}
+n^2\;e_{z2}e^{-i\beta \left(a-b\right)},\lbl{NormalDBc1.1}\\
&\nonumber\\
&\vb{\hat k}\cross\vb{e}_{1}e^{i\beta\left(a+b\right)}
+\vb{\hat k}\cross\vb{e}_{2}e^{-i\beta\left(a+b\right)}\nonumber\\
&=\vb{\hat k}\cross\vb{e}_{1}^{-}e^{i\beta^{-} \left(a+b-L_z\right)}
+\vb{\hat k}\cross\vb{e}_{2}^{-}e^{-i\beta^{-} \left(a+b-L_z\right)},\lbl{ParallelEBc2.1}\\
&\nonumber\\
&n^2\;e_{z1}e^{i\beta\left(a+b\right)}
+n^2\;e_{z2}e^{-i\beta\left(a+b\right)}\nonumber\\
&=n_0^2\;e_{z1}^{-}e^{i\beta^{-} \left(a+b-L_z\right)}
+n_0^2\;e_{z2}^{-}e^{-i\beta^{-} \left(a+b-L_z\right)}.\lbl{NormalDBc2.1}
\end{align}

\noindent Let us first consider the case of TE modes. For these modes $e_{iz}=e_{iz}^{-}=0$ for $i=1,2$. Therefore, for TE modes equations \rf{NormalDBc1.1} and \rf{NormalDBc2.1} are automatically satisfied.

For a given transverse wave number $\xi$, introduce a positively oriented orthonormal triad $\left(\vb{\hat\xi},\vb{\hat\eta},\vb{\hat k}\right)$ where $\vb{\hat\xi}$ is a unit vector pointing along the transverse wave number vector $\xi$, $\vb{\hat \eta}=\vb{\hat k}\cross\vb{\hat \xi}$ and $\vb{\hat k}$ is a unit vector along the $z$ axis.

Then, using the orthonormal triad to define the scalar quantities $\xi,e^{-}_{1,2}$ and $e_{1,2}$
\begin{align}
\vb{\xi}&=\xi\vb{\hat \xi},\\
\vb{e}^{-}_{1,2}&=e^{-}_{1,2}\vb{\hat \eta},\\
\vb{e}_{1,2}&=e_{1,2}\vb{\hat \eta},
\end{align}
we have for $i=1,2$
\begin{align}
\vb{\hat k}\cross\vb{e}^{-}_i&=\vb{\hat k}\cross\left(e_i^{-}\vb{\hat\eta}\right)=-e^{-}_i\vb{\hat \xi}.\nonumber
\end{align}
In a similar way we get
\begin{align}
\vb{\hat k}\cross\vb{e}_i&=-e_i\vb{\hat \xi}.\nonumber
\end{align}
Using these expressions in \rf{ParallelEBc1.1} and \rf{ParallelEBc2.1} give us the simplifies equaations
\begin{align}
&e_{1}^{-}e^{i\beta^{-}\left(a-b\right)}
+e_{2}^{-}e^{-i\beta^{-}\left(a-b\right)}\nonumber\\
&=e_{1}e^{i\beta \left(a-b\right)}
+e_{2}e^{-i\beta \left(a-b\right)},\lbl{ParallelEBc1.2}\\
&\nonumber\\
&e_{1}e^{i\beta\left(a+b\right)}
+e_{2}e^{-i\beta\left(a+b\right)}\nonumber\\
&=e_{1}^{-}e^{i\beta^{-} \left(a+b-L_z\right)}
+e_{2}^{-}e^{-i\beta^{-} \left(a+b-L_z\right)}.\lbl{ParallelEBc2.2}
\end{align}
We also have for TE modes
\begin{align}
\left(\vb{\xi},\beta^{-}\right)\cross\vb{E}_{1}^{-}&=\left(\xi\vb{\hat\xi}+\beta^{-}\vb{\hat k}\right)\cross\left(e_1^{-}\vb{\hat \eta}\right)
=e_1^{-}\xi\;\vb{\hat k}-\beta^{-}e_1^{-}\vb{\hat\xi},
\end{align}
and similar expressions for the other cross products in the magnetic boundary conditions. Using these expressions for the cross products, the magnetic boundary condition at $z=a-b$, \rf{MagBc1}, can be written in the form
\begin{align}
&\left(e_1^{-}\xi\;\vb{\hat k}-\beta^{-}e_1^{-}\vb{\hat\xi}\right)e^{i\beta^{-}\left(a-b\right)}
+\left(e_2^{-}\xi\;\vb{\hat k}+\beta^{-}e_2^{-}\vb{\hat\xi}\right)e^{-i\beta^{-}\left(a-b\right)}\nonumber\\
&=\left(e_1\xi\;\vb{\hat k}-\beta e_1\vb{\hat\xi}\right)e^{i\beta \left(a-b\right)}
+\left(e_2\xi\;\vb{\hat k}+\beta e_2\vb{\hat\xi}\right)e^{-i\beta \left(a-b\right)}.\lbl{MagBc1.1}
\end{align}
This single vector equation splits into two scalar equations
\begin{align}
&e_1^{-}e^{i\beta^{-}\left(a-b\right)}
+e_2^{-} e^{-i\beta^{-}\left(a-b\right)}\nonumber\\
&=e_1 e^{i\beta \left(a-b\right)}
+e_2 e^{-i\beta \left(a-b\right)}.\lbl{MagBc1.2}
\end{align}
and
\begin{align}
&-\beta^{-}e_1^{-}e^{i\beta^{-}\left(a-b\right)}
+\beta^{-}e_2^{-}e^{-i\beta^{-}\left(a-b\right)}\nonumber\\
&=-\beta e_1e^{i\beta \left(a-b\right)}
+\beta e_2e^{-i\beta \left(a-b\right)}.\lbl{MagBc1.3}
\end{align}
Repeating these calculations for the magnetic boundary condition at $z=a+b$ we get the two scalar scalar equations 
\begin{align}
&e_1 e^{i\beta\left(a+b\right)}
+e_2 e^{-i\beta\left(a+b\right)}\nonumber\\
&=e_1^{-} e^{i\beta^{-} \left(a+b-L_z\right)}
+e_2^{-} e^{-i\beta^{-} \left(a+b-L_z\right)}.\lbl{MagBc1.4}
\end{align}
and
\begin{align}
&-\beta e_1e^{i\beta\left(a+b\right)}
+\beta e_2e^{-i\beta\left(a+b\right)}\nonumber\\
&=-\beta^{-} e^{-}_1e^{i\beta^{-} \left(a+b-L_z\right)}
+\beta^{-} e_2^{-}e^{-i\beta^{-} \left(a+b-L_z\right)}.\lbl{MagBc1.5}
\end{align}
Observe that equations \rf{ParallelEBc1.2} and \rf{MagBc1.2} are identical, as are equations \rf{ParallelEBc2.2} and \rf{MagBc1.4}. Thus we have four remaining equations \rf{MagBc1.2},\rf{MagBc1.3},\rf{MagBc1.4} and \rf{MagBc1.5} for the four unknown quantities $e_1,e_2,e_1^{-},e_2^{-}$. The four equations are 
\begin{align}
&e_1^{-}e^{i\beta^{-}\left(a-b\right)}
+e_2^{-} e^{-i\beta^{-}\left(a-b\right)}\nonumber\\
&=e_1 e^{i\beta \left(a-b\right)}
+e_2 e^{-i\beta \left(a-b\right)},\lbl{TeBc1}\\
&\nonumber\\
&-\beta^{-}e_1^{-}e^{i\beta^{-}\left(a-b\right)}
+\beta^{-}e_2^{-}e^{-i\beta^{-}\left(a-b\right)}\nonumber\\
&=-\beta e_1e^{i\beta \left(a-b\right)}
+\beta e_2e^{-i\beta \left(a-b\right)},\lbl{TeBc2}\\
&\nonumber\\
&e_1 e^{i\beta\left(a+b\right)}
+e_2 e^{-i\beta\left(a+b\right)}\nonumber\\
&=e_1^{-} e^{i\beta^{-} \left(a+b-L_z\right)}
+e_2^{-} e^{-i\beta^{-} \left(a+b-L_z\right)},\lbl{TeBc3}\\
&\nonumber\\
&-\beta e_1e^{i\beta\left(a+b\right)}
+\beta e_2e^{-i\beta\left(a+b\right)}\nonumber\\
&=-\beta^{-} e^{-}_1e^{i\beta^{-} \left(a+b-L_z\right)}
+\beta^{-} e_2^{-}e^{-i\beta^{-} \left(a+b-L_z\right)},\lbl{TeBc4}
\end{align}
which can be written as the following two matrix equations 
\begin{align}
\left(
\begin{array}{ccc}
\alpha&\alpha^{-1} & \\
\beta^{-}\alpha&-\beta^{-}\alpha^{-1} & \\
\end{array}
\right)\left(
\begin{array}{cc}
e_1^{-}&\\
e_2^{-}&\\
\end{array}
\right)&=\left(
\begin{array}{ccc}
\gamma&\gamma^{-1} & \\
\beta\gamma&-\beta\gamma^{-1} & \\
\end{array}
\right)\left(
\begin{array}{cc}
e_1&\\
e_2&\\
\end{array}
\right),\nonumber\\
&\nonumber\\
\left(
\begin{array}{ccc}
\delta&\delta^{-1} & \\
\beta\delta&-\beta\delta^{-1} & \\
\end{array}
\right)\left(
\begin{array}{cc}
e_1&\\
e_2&\\
\end{array}
\right)&=\left(
\begin{array}{ccc}
\omega&\omega^{-1} & \\
\beta^{-}\omega&-\beta^{-}\omega^{-1} & \\
\end{array}
\right)\left(
\begin{array}{cc}
e_1^{-}&\\
e_2^{-}&\\
\end{array}
\right),\nonumber
\end{align}
where we have defined 
\begin{align}
\alpha&=e^{i\beta^{-}\left(a-b\right)}\nonumber\\
\gamma&=e^{i\beta\left(a-b\right)}\nonumber\\
\delta&=e^{i\beta\left(a+b\right)}\nonumber\\
\omega&=e^{i\beta^{-}\left(a+b-L_z\right)}\lbl{Phasesdefined}
\end{align}
We can now easily eliminate the variables $e_1^{-}$ and $e_2^{-}$ and get the following single matric equation for the remaining two variables $e_1$ and $e_2$
\begin{align}
M_{TE}\left(
\begin{array}{cc}
e_1&\\
e_2&\\
\end{array}
\right)=0,
\end{align}
where 
\begin{align}
M_{TE}&=\left(
\begin{array}{ccc}
\delta&\delta^{-1} & \\
\beta\delta&-\beta\delta^{-1} & \\
\end{array}
\right)\nonumber\\
&-\left(
\begin{array}{ccc}
\omega&\omega^{-1} & \\
\beta^{-}\omega&-\beta^{-}\omega^{-1} & \\
\end{array}
\right)\left(
\begin{array}{ccc}
\alpha&\alpha^{-1} & \\
\beta^{-}\alpha&-\beta^{-}\alpha^{-1} & \\
\end{array}
\right)^{-1}\left(
\begin{array}{ccc}
\gamma&\gamma^{-1} & \\
\beta\gamma&-\beta\gamma^{-1} & \\
\end{array}
\right).
\end{align}
The eigenvalue equation for TE modes is then evidently given by 
\begin{align}
Det\left(M_{TE}\right)=0.\lbl{EigenvalueEquationTEModes}
\end{align}
It is useful to split the set of TE modes into two types. The first type corresponds to both $\beta$ and $\beta^{-}$ being real. We will call this type  a {\it scattering} mode. The second type corresponds to $\beta$ being real but $\beta^{-}$ being imaginary.  We will call this type of modes for {\it bound}. Note that in  the eigenvalue equation for bound modes \rf{BoundEigenvalueEquationTEModes} we have redefined $\beta^{-}$ to be the imaginary part of the original $\beta^{-}$. Explicitly we have in \rf{BoundEigenvalueEquationTEModes}
\begin{align}
 \beta^{-}=\sqrt{\xi^2-\left(\frac{\omega}{c}\right)^2n_0^2}.
 \end{align}
  From \rf{EigenvalueEquationTEModes} we find that the explicit eigenvalue equations for scattering and bound TE modes are
\begin{align}
&\left(\beta+\beta^{-}\right)^2\cos\left(L\left(\beta-\beta^{-}\right)+\beta^{-}L_z\right)\nonumber\\
&-\left(\beta-\beta^{-}\right)^2\cos\left(L\left(\beta+\beta^{-}\right)-\beta^{-}L_z\right)=4\beta\beta^{-},\lbl{FreeEigenvalueEquationTEModes}\\
&\nonumber\\
&2\beta\beta^{-}\cosh\left(\beta^{-}\left(L-L_z\right)\right)\cos\left(\beta L\right)\nonumber\\
&+\left(\beta^2-\left(\beta^{-}\right)^2\right)\sinh\left(\beta^{-}\left(L-L_z\right)\right)\sin\left(\beta L\right)=2\beta\beta^{-},\lbl{BoundEigenvalueEquationTEModes}
\end{align}
where now $L=2b$ denote the thickness of the slab.
 
For the case of TM modes we get in an entirely similar way a matrix equation
\begin{align}
M_{TM}\left(
\begin{array}{cc}
e_1&\\
e_2&\\
\end{array}
\right)=0,
\end{align}
where now
\begin{align}
M_{TM}&=\left(
\begin{array}{ccc}
\delta&\delta^{-1} & \\
\frac{n^2}{\beta}\delta&-\frac{n^2}{\beta}\delta^{-1} & \\
\end{array}
\right)\nonumber\\
&-\left(
\begin{array}{ccc}
\omega&\omega^{-1} & \\
\frac{n_0^2}{\beta^{-}}\omega&-\frac{n_0^2}{\beta^{-}}\omega^{-1} & \\
\end{array}
\right)\left(
\begin{array}{ccc}
\alpha&\alpha^{-1} & \\
\frac{n_0^2}{\beta^{-}}\alpha&-\frac{n_0^2}{\beta^{-}}\alpha^{-1} & \\
\end{array}
\right)^{-1}\left(
\begin{array}{ccc}
\gamma&\gamma^{-1} & \\
\frac{n^2}{\beta}\gamma&-\frac{n^2}{\beta}\gamma^{-1} & \\
\end{array}
\right).
\end{align}
Thus, the eigenvalue equation for TM modes is 
\begin{align}
Det\left(M_{TM}\right)=0\lbl{EigenvalueEquationTMModes}.
\end{align}
The two variables $e_1$ and $e_2$ determine the second pair of unknown variables $e_1^{-}$ and $e_2^{-}$ using the matrix equation
\begin{align}
\left(
\begin{array}{cc}
e_1^{-}&\\
e_2^{-}&\\
\end{array}
\right)&=\left(
\begin{array}{ccc}
\alpha&\alpha^{-1} & \\
\frac{n_0^2}{\beta^{-}}\alpha&-\frac{n_0^2}{\beta^{-}}\alpha^{-1} & \\
\end{array}
\right)^{-1}\left(
\begin{array}{ccc}
\gamma&\gamma^{-1} & \\
\frac{n^2}{\beta}\gamma&-\frac{n^2}{\beta}\gamma^{-1} & \\
\end{array}
\right)\left(
\begin{array}{cc}
e_1&\\
e_2&\\
\end{array}
\right),\nonumber
\end{align}
and the four quantities  $e_1$, $e_2$, $e^{-}_1$, $e^{-}_2$ determine the electric field amplitudes $e_{iz}$,$\vb{e}_i$,$e_{iz}^{-}$, $\vb{e}_{i}^{-}$ for $i=1,2$, using the identities
\begin{align}
e_{1z}&=-\frac{\xi}{\beta}e_1,\nonumber\\
e_{2z}&=\frac{\xi}{\beta}e_2,\nonumber\\
e^{-}_{1z}&=-\frac{\xi}{\beta^{-}}e_1^{-},\nonumber\\
e^{-}_{2z}&=\frac{\xi}{\beta^{-}}e_2^{-},\nonumber\\
\vb{e}_1&=e_1\vb{\hat\xi},\nonumber\\
\vb{e}_2&=e_2\vb{\hat\xi},\nonumber\\
\vb{e}^{-}_1&=e^{-}_1\vb{\hat\xi},\nonumber\\
\vb{e}^{-}_2&=e^{-}_1\vb{\hat\xi}.\nonumber\\
\end{align}
 From \rf{EigenvalueEquationTMModes} we find that the explicit eigenvalue equations for scattering and bound TM modes are
\begin{align}
&\left(n^2\beta+n_0^2\beta^{-}\right)^2\cos\left(L\left(\beta-\beta^{-}\right)+\beta^{-}L_z\right)\nonumber\\
&-\left(n^2\beta-n_0^2\beta^{-}\right)^2\cos\left(L\left(\beta+\beta^{-}\right)-\beta^{-}L_z\right)=4n^2n_0^2\beta\beta^{-},\lbl{FreeEigenvalueEquationTMModes}\\
&\nonumber\\
&2n^2n_0^2\beta\beta^{-}\cosh\left(\beta^{-}\left(L-L_z\right)\right)\cos\left(\beta L\right)\nonumber\\
&+\left(n_0^4\beta^2-n^4\left(\beta^{-}\right)^2\right)\sinh\left(\beta^{-}\left(L-L_z\right)\right)\sin\left(\beta L\right)=2n^2n_0^2\beta\beta^{-},\lbl{FreeEigenvalueEquationTMModes}
\end{align}

\paragraph{Eigenvalue equations for scattering modes  in the large cavity limit} 

The eigenvalue equation \rf{FreeEigenvalueEquationTEModes} for TE scattering modes  can be rewritten into the form
\begin{align}
\alpha\cos(L_z\beta^{-})+\gamma\sin(L_z\beta^{-})=\delta,\lbl{TEEquation1}
\end{align}
where we have defined
\begin{align}
\alpha&=(\beta+\beta^{-})^2\cos(L(\beta-\beta^{-}))-(\beta-\beta^{-})^2\cos(L(\beta+\beta^{-})),\lbl{alpha1}\\
\gamma&=(\beta+\beta^{-})^2\sin(L(\beta-\beta^{-}))+(\beta-\beta^{-})^2\sin(L(\beta+\beta^{-})),\lbl{gamma1}\\
\delta&=4\beta\beta^{-}.\lbl{delta1}
\end{align}
The goal is now to write down a formal solution to the eigenvalue equation \rf{TEEquation1} and for that purpose we introduce a phase angle $\phi$ as the solution to the equation
\begin{align}
\tan(\phi)=\frac{\gamma}{\alpha}.\lbl{phi1}
\end{align}
The general solution to \rf{phi1} is
\begin{align}
\phi_n=\arctan\left(\frac{\gamma}{\alpha}\right)+n\pi,
\end{align}
where $\arctan$ is the standard branch of the inverse tangent function and where $n$ is an arbitrary integer that represents all the other branches. Using standard trigonometric formulas, the eigenvalue equation \rf{TEEquation1} can be rewritten into the form
\begin{align}
\cos(L_z\beta^{-}+\phi+n\pi)=\frac{\delta\cos(\phi)}{\alpha},\lbl{TEEquation2}
\end{align}
where we now have defined $\phi=\arctan\left(\frac{\gamma}{\alpha}\right)$.  Because of the $2\pi$ periodicity of $\cos$ only the two branches $n=0,1$ produce different eigenvalue equations \rf{TEEquation2}. Let us introduce the symbol $s=0,1$ to represent the two possibilities. The variable $s$ will be calles the {\it spin} variable and our eigenvalue equation is now of the form
\begin{align}
\cos(L_z\beta^{-}+\phi+s\pi)=\frac{\delta\cos(\phi)}{\alpha},\lbl{TEEquation2}
\end{align}

Since
\begin{align}
\beta^{-}=\sqrt{\left(\frac{\omega}{c}\right)^2n^2_0-\xi^2},\nonumber
\end{align}
 equation \rf{TEEquation2} can be formally solved with respect to $\omega$ and we get 
\begin{align}
\omega=\left(\frac{c}{n_0}\right)\left(\xi^2+\left[k_z+\frac{1}{L_z}\left\{\arccos\left(\frac{\delta\cos(\phi)}{\alpha}\right)-\phi -s\pi\right\}\right]^2\right)^{\frac{1}{2}},\lbl{omega1}
\end{align}
where $\xi$ and $k_z$ are discrete variables
\begin{align}
\xi&=\left(\left(\frac{2\pi n_x}{L_x}\right)^2+\left(\frac{2\pi n_y}{L_y}\right)^2\right)^{\frac{1}{2}}\lbl{ksi}\\
k_z&=\frac{2\pi n_z}{L_z},\lbl{kz1}
\end{align}
 and where $\arccos$ is the standard branch of the inverse cosine function. The branch multiplicity has been absorbed into the discrete variable $k_z$.
 
We now introduce the large cavity limit that consists of letting $L_x,L_y,L_z\rightarrow\infty$ in such a way that $\xi$ and $k_z$ become continuous, real valued variables. In the large cavity limit equation \rf{omega1} can be written in the form
\begin{align}
\omega=\left(\frac{c}{n_0}\right)\left(\xi^2+k_z^2\right)^{\frac{1}{2}}+\frac{1}{L_z}\left(\frac{c}{n_0}\right)\frac{k_z\left(\arccos\left(\frac{\delta\cos(\phi)}{\alpha}\right)-\phi-s\pi\right)}{\left(\xi^2+k_z^2\right)^{\frac{1}{2}}},\lbl{omega2}
\end{align}
We find a formula for the eigenvalues of TE scattering modes correct to first order in the large-cavity parameter $\frac{1}{L_z}$ by iterating the expression \rf{omega2} once. This gives us two solutions families $\omega_s$, each parametrized by the two continuous variables $\xi$ and $k_z$
\begin{align}
\omega_s(\xi,k_z)=\left(\frac{c}{n_0}\right)\left(\xi^2+k_z^2\right)^{\frac{1}{2}}+\frac{1}{L_z}h_s(\xi,k_z),\lbl{omega21}
\end{align}
where 
\begin{align}
h_s(\xi,k_z)=\left(\left(\frac{c}{n_0}\right)\frac{k_z\left(\arccos\left(\frac{\delta\cos(\phi)}{\alpha}\right)-\phi-s\pi\right)}{\left(\xi^2+k_z^2\right)^{\frac{1}{2}}}\right)_{\omega=\left(\frac{c}{n_0}\right)\left(\xi^2+k_z^2\right)^{\frac{1}{2}}}.\lbl{h1}
\end{align}
\rf{omega21} with $h$ given by \rf{h1} is the general formula for the first correction to the eigenvalues of TE scattering modes. We find the exact same formulas for the eigenvalues of the TM scattering modes.

\paragraph{Eigenvalue equations for bound states  in the large cavity limit} 
In the large cavity limit the eigenvalue equation for bound TE modes is up to exponentially small terms given by 
\begin{align}
\tan(\beta L)=\frac{2\beta\beta^-}{(\beta^2-(\beta^-)^2)^\frac{1}{2}}.\lbl{LargeCavityTEBound2}
\end{align}
It is easy to see that this equation has a finite number of solutions which we enumerate with a parameter $j$ 
\begin{align}
0<j<J(\xi)=\xi\left(\frac{L}{\pi}\right)\frac{\sqrt{n^2-n_0^2}}{n_0}.\lbl{BoundStateUpperBound}
\end{align}
We find the exact same equation for the eigenvalues of  bound TM  modes.

\paragraph{Approximations for mode sums in term of mode integrals in the large cavity limit}
In the  large cavity limit for TE and TM scattering modes discussed in the previous section,  two separate discrete variables  became  continuous in the limit. 
\begin{align}
\xi&=\left(\left(\frac{2\pi n_x}{L_x}\right)^2+\left(\frac{2\pi n_y}{L_y}\right)^2\right)^{\frac{1}{2}}\lbl{ksi2}\\
k_z&=\frac{2\pi n_z}{L_z},\lbl{kz2}
\end{align}
The first discrete  variable $\xi$ comes from the periodicity requirement for the cavity fields in the transverse $xy$ directions, and the second discrete variable $k_z$ comes from the enumeration of the infinite set of branches for the inverse cosine function.
The area of the 2D transverse wave number space taken up by a single discrete $\xi-$mode is
\begin{align}
\delta_{\xi}=\frac{4\pi^2}{L_xL_y}=\frac{4\pi^2}{A},\lbl{deltaksi}
\end{align}
where $A=L_xL_y$ is the surface are of the slab, and the length in the 1D longitudinal wave number space  taken up by a single discrete $k_z-$mode is
\begin{align}
\delta_{k_z}=\frac{2\pi}{L_z}.\lbl{deltakz}
\end{align}
On the other hand, the area taken up in the 2D transverse wave number space of modes with wave numbers between $\xi$ and $\xi+d\xi$ is
$2\pi\xi d\xi$, and the lengt  taken up in 1d longitudinal wave number space of modes with wave number between $k_z$ and $k_z+dk_z$ is simply $dk_z$. Therefore, the mode density in $(\xi,k_z)$ space, i. e. the number of cavity modes per unit volume in 3D $(\xi,k_z)$ space, is
\begin{align}
D_s(\xi,k_z)&=2\;\left(\frac{2\pi\xi d\xi}{\delta_{\xi}}\right)\left(\frac{dk_z}{\delta_{k_z}}\right)\nonumber\\
&=\left(\frac{V_0\xi}{2\pi^2}\right)\;d\xi dk_z,\lbl{ModeDensityScatteringModes}
\end{align}
where $V_0=L_xL_yL_z$ is the volume of the cavity. The factor of two appearing in the definition of the mode density is there to take into account that there are both TE and TM modes.

In the large cavity limit for the TE and TM bound modes only one discrete variable became continuous in the limit
\begin{align}
\xi&=\left(\left(\frac{2\pi n_x}{L_x}\right)^2+\left(\frac{2\pi n_y}{L_y}\right)^2\right)^{\frac{1}{2}},
\end{align}
the other variable, now enumerating the finite set of bound modes stays discrete.
Repeating the argument just given for the scattering case now gives us the mode density
\begin{align}
D_b(\xi)=\left(\frac{A\xi}{\pi}\right)\;d\xi.\lbl{ModeDensityBoundModes}
\end{align}
We can thus conclude that for any function $f(\omega_{\vb{m}})$ we can approximate mode sums with mode integrals using 
\begin{align}
\sum_{\vb{m}}f(\omega_{\vb{m}})&\approx\frac{V_0}{2\pi^2}\sum_s\int_0^{\infty}d\xi\;\xi\int_{-\infty}^{\infty}dk_zf(\omega_s(\xi,k_z))\nonumber\\
&+\frac{A}{\pi}\int_0^{\infty}d\xi\;\xi\sum_{j=0}^{J(\xi)}f(\omega_j(\xi)),\lbl{ModeSumApproximation}
\end{align}
where
\begin{align}
J(\xi)=\xi\left(\frac{L}{\pi}\right)\frac{\sqrt{n^2-n_0^2}}{n_0},\lbl{BoundStateUpperBound2}
\end{align}
and where $\omega_s(\xi,k_z)$ are the two families of scattering eigenvalue as a function of the continuous variables $\xi$ and $k_z$ from \rf{omega21} and $\omega_j(\xi)$ is the finite set of bound mode eigenvalues as a function of $\xi$ that are the solutions of \rf{LargeCavityTEBound2} .

\noindent All thermodynamical quantities can be expressed in terms of mode sums and we now have the tools to convert these mode sums into mode integrals and thereby facilitate the calculation of pressure, particle number, total energy etc. The mode integrals for quantities of interest will be complicated, this is evident from formula \rf{ModeSumApproximation}, and  must almost certainly  be done purely  numerically or approximated using a combination of numerical and semi analytic asymptotic methods. This is however a story for another day.

\section*{Acknowledgements}
The author is thankful to Miroslav Kolesik and Masud Manpurisur, both faculty at the College of Optical Sciences at the  University of Arizona,  for inspiration and support while developing these notes. The author would also like to thank his friend and college Valentin V. Lychagin, professor emeritus, at the department of mathematics, the Arctic University Of Norway, Troms\o\; Norway. It was through Valentin that the author came to be familiar with the information theoretic foundation for thermodynamics and all that  it entails. The author is also thankful for  support from the department of mathematics and statistics at the Arctic University of Norway, from  the Arizona Center for Mathematical Sciences at the University of Arizona and for the support from the Air Force Office for Scientific Research under grant \# FA9550-16-1-0088.

\section{Appendix A}
From section 3 we have the following expression for the electromagnetic field in the cavity
 \begin{align}
  \vb{E}(\vb{x},t)&=-\sum_{\vb{m}}\frac{\dot q_{\vb{m}}(t)}{\sqrt{\epsilon_0}}\;\vb{u}_{\vb{m}}(\vb{x}),\nonumber\\
  \vb{H}(\vb{x},t)&=\sum_{\vb{m}}\frac{q_{\vb{m}}(t)}{\mu_0\sqrt{\epsilon_0}}\;\curl \vb{u}_{\vb{m}}(\vb{x}),\lbl{A.1}
  \end{align}
  The total energy in the cavity is 
  \begin{align}
  \mathcal{H}=\mathcal{H}_E+\mathcal{H}_H\lbl{A.2},
  \end{align}
  where 
  \begin{align}
  \mathcal{H}_E&=\frac{1}{2}\int_{D}d\vb{x}\;\epsilon_0n^2(\vb{x})\vb{E}(\vb{x})\cdot\vb{E}(\vb{x}),\lbl{A.3}\\
   \mathcal{H}_H&=\frac{1}{2}\int_{D}d\vb{x}\;\mu_0\vb{H}(\vb{x})\cdot\vb{H}(\vb{x})\lbl{A.4}.
  \end{align}
  For the electric part of the energy we have using \rf{A.1} and \rf{A.3}
  \begin{align}
  \mathcal{H}_E&=\frac{1}{2}\int_{D}d\vb{x}\;\epsilon_0n^2(\vb{x})\vb{E}(\vb{x})\cdot\vb{E}(\vb{x})\nonumber\\
  &=\frac{1}{2}\sum_{\vb{m}\vb{m}'}q'_{\vb{m}}q'_{\vb{m}'}\int_{D}d\vb{x}\;n^2(\vb{x})\vb{u}_{\vb{m}}(\vb{x})\cdot \vb{u}_{\vb{m}'}(\vb{x})\nonumber\\
  &=\frac{1}{2}\sum_{\vb{m}}\;q^{`2}_{\vb{m}}.\lbl{A.5}
  \end{align}
  For the magnetic part we have
  \begin{align}
   \mathcal{H}_H&=\frac{1}{2}\int_{D}d\vb{x}\;\mu_0(\vb{x})\vb{H}(\vb{x})\cdot\vb{H}(\vb{x})\nonumber\\
  &=\frac{1}{2}\sum_{\vb{m}\vb{m}'}\frac{1}{\mu_0\epsilon_0}q_{\vb{m}}q_{\vb{m}'}\alpha_{\vb{m}\vb{m}'},\lbl{A.6}
  \end{align} 
  where 
 \begin{align}
 \alpha_{\vb{m}\vb{m}'} &=\int_{D}d\vb{x}\;\curl{\vb{u}_{\vb{m}}(\vb{x})}\cdot\curl{ \vb{u}_{\vb{m}'}(\vb{x})}.\lbl{A.7}
 \end{align}
 Using the fundamental integral identity
 \begin{align}
 \div{(\vb{a}\cross\vb{b})}=\vb{b}\cdot(\curl{\vb{a}})-\vb{a}\cdot(\curl{\vb{b}}),\nonumber
 \end{align}
 and the geometry of the situation, \rf{A.7} can be written in the form
 \begin{align}
  \alpha_{\vb{m}\vb{m}'}&= \int_{D}d\vb{x}\;\curl{\vb{u}_{\vb{m}}(\vb{x})}\cdot\curl{ \vb{u}_{\vb{m}'}(\vb{x})}\nonumber\\
  &+\sum_{j=1}^p \int_{D_j}d\vb{x}\;\curl{\vb{u}_{\vb{m}}(\vb{x})}\cdot\curl{ \vb{u}_{\vb{m}'}(\vb{x})}.\lbl{A.8}
 \end{align}
 For any $j$ and vector field $\vb{a}$ defined inside and outside of $D_j$,we introduce the notation
 \begin{align}
 \vb{a}^{j(+)}(\vb{x})=\lim_{
{\small  \begin{array}{ccc}
\vb{y}\rightarrow\vb{x} & \\
\vb{y}\in D_j^{c} & 
\end{array} }}\vb{a}(\vb{y})\nonumber,\\
 \vb{a}^{j(-)}(\vb{x})=\lim_{
{\small  \begin{array}{ccc}
\vb{y}\rightarrow\vb{x} & \\
\vb{y}\in D_j & 
\end{array} }}\vb{a}(\vb{y})\lbl{A.9},
\end{align}
where $D_j^{c}$ denotes the complement of $D_j$. Using the notation \rf{A.9} in \rf{A.8} we have

\begin{align}
\alpha_{\vb{m}\vb{m}'}&=\int_{D_0} d\vb{x}\left\{\div\left(\vb{u}_{\vb{m}}\cross\left(\curl{\vb{u}_{\vb{m}'}}\right)\right)+\vb{u}_{\vb{m}}\cdot\curl{\left(\curl{\vb{u}_{\vb{m}'}}\right)}\right\}\nonumber\\
&+\sum_{j=1}^p\int_{D_j} d\vb{x}\left\{\div\left(\vb{u}_{\vb{m}}\cross\left(\curl{\vb{u}_{\vb{m}'}}\right)\right)+\vb{u}_{\vb{m}}\cdot\curl{\left(\curl{\vb{u}_{\vb{m}'}}\right)}\right\}\nonumber\\
&=\int_{\Gamma_0} d\vb{x}\;\vb{n}\cdot\left(\vb{u}^{0(-)}_{\vb{m}}\cross\left(\curl{\vb{u}^{0(-)}_{\vb{m}'}}\right)\right)\nonumber\\
&-\sum_{j=1}^p\int_{\Gamma_j} d\vb{x}\;\vb{n}\cdot\left\{\left(\vb{u}^{j(-)}_{\vb{m}}\cross\left(\curl{\vb{u}^{j(-)}_{\vb{m}'}}\right)\right)-\left(\vb{u}^{j(+)}_{\vb{m}}\cross\left(\curl{\vb{u}^{j(+)}_{\vb{m}'}}\right)\right)\right\}
\nonumber\\
&+\sum_{j=0}^p\int_{D_j} d\vb{x}\;\vb{u}_{\vb{m}}\cdot\curl{\left(\curl{\vb{u}_{\vb{m}'}}\right)}\nonumber\\
&=\int_{\Gamma_0} d\vb{x}\;\vb{n}\cdot\left(\vb{u}^{0(-)}_{\vb{m}}\cross\left(\curl{\vb{u}^{0(-)}_{\vb{m}'}}\right)\right)+\sum_{j=0}^p\int_{D_j} d\vb{x}\;\left(\frac{\omega_{\vb{m'}}}{c}\right)^2\;n_j^2\vb{u}_{\vb{m}}\cdot\vb{u}_{\vb{m}'},\nonumber
\end{align}
where we have used the boundary conditions on the internal boundaries. If we know use the boundary conditions on the outer boundary of the cavity we finally get
\begin{align}
\alpha_{\vb{m}\vb{m}'}&=\left(\frac{\omega_{\vb{m}}}{c}\right)^2\int_{D} d\vb{x}\;\;n(\vb{x})^2\vb{u}_{\vb{m}}\cdot\vb{u}_{\vb{m}'}\nonumber.
\end{align}
Given this, \rf{A.6} now implies that 
\begin{align}
 \mathcal{H}_H&=\frac{1}{2}\sum_{\vb{m}}\omega_{\vb{m}}^2q_{\vb{m}}^2,\nonumber
 \end{align}
 and we finally get
 \begin{align}
 \mathcal{H}=\frac{1}{2}\sum_{\vb{m}}\left(q^{`2}_{\vb{m}}+\omega_{\vb{m}}^2q_{\vb{m}}^2\right).
 \end{align}

\bibliographystyle{plain}
\bibliography{ThermodynamicsOfLightRef}

\end{document}